\documentclass[twoside,journey]{IEEEtran}
\usepackage{makecell}
\usepackage{hyperref}
\usepackage{array}
\usepackage{graphicx,amssymb,amsmath}
\usepackage{multicol}
\usepackage[noadjust]{cite}
\usepackage{setspace}
\usepackage{subfigure}
\usepackage{graphicx}
\usepackage{float}
\usepackage {url}
\usepackage{stfloats}
\usepackage{amsthm,pifont}
\usepackage{flushend}
\usepackage{cases,subeqnarray}
\usepackage{bm,multirow,bigstrut}
\usepackage{amsmath, amsthm, amssymb}
\usepackage{textcomp}
\usepackage{latexsym,bm}
\usepackage{booktabs}
\usepackage{xcolor}
\usepackage{mathtools}
\usepackage{dsfont}
\usepackage{extarrows}
\usepackage{epsfig}
\usepackage{epsfig}
\usepackage{epstopdf}
\usepackage[noend]{algpseudocode}
\usepackage{algorithmicx,algorithm}
\usepackage{makecell}
\usepackage{colortbl}
\usepackage{booktabs}
\usepackage{graphicx} 
\usepackage{array}    
\usepackage{hyperref} 
\usepackage{booktabs}
\usepackage{enumitem}
\usepackage{diagbox}
\usepackage{tikz}
\usetikzlibrary{trees,shadows.blur}
\usepackage{forest}
\theoremstyle{plain}

\theoremstyle{plain}

\usepackage{amsmath}

\newcommand{\ignore}[1]{{{\color{yellow} }}}
\renewcommand{\arraystretch}{1.5} 
\definecolor{blue-green}{rgb}{0.0, 0.87, 0.87}
\IEEEoverridecommandlockouts
\begin{document}

\title{Generative AI for Secure Physical Layer Communications: A Survey}
\author{Changyuan Zhao, Hongyang Du, Dusit Niyato, \textit{Fellow}, IEEE, Jiawen Kang, Zehui Xiong, Dong In Kim, \textit{Fellow}, IEEE, Xuemin (Sherman) Shen, \textit{Fellow}, IEEE, and Khaled B. Letaief, \textit{Fellow}, IEEE
\thanks{C. Zhao, H. Du, and D. Niyato are with the School of Computer Science and Engineering, Nanyang Technological University, Singapore (e-mail: zhao0441@e.ntu.edu.sg; hongyang001@e.ntu.edu.sg; dniyato@ntu.edu.sg).}
\thanks{J. Kang is with the School of Automation, Guangdong University of Technology, China. (e-mail: kavinkang@gdut.edu.cn).
}
\thanks{Z. Xiong is with the Pillar of Information Systems Technology and Design, Singapore University of Technology and Design, Singapore (e-mail: zehui xiong@sutd.edu.sg).}
\thanks{D. I. Kim is with the Department of Electrical and Computer Engineering, Sungkyunkwan University, Suwon 16419, South Korea (email:dikim@skku.ac.kr).}
\thanks{X. Shen is with the Department of Electrical and Computer Engineering, University of Waterloo, Canada (e-mail: sshen@uwaterloo.ca).
}
\thanks{Khaled B. Letaief is with the Department of Electrical and Computer Engineering, Hong Kong University of Science and Technology, Hong Kong (e-mail: eekhaled@ust.hk).
}
}

\maketitle
\vspace{-1cm}

\begin{abstract}

Generative Artificial Intelligence (GAI) stands at the forefront of AI innovation, demonstrating rapid advancement and unparalleled proficiency in generating diverse content. Beyond content creation, GAI has significant analytical abilities to learn complex data distribution, offering numerous opportunities to resolve security issues. In the realm of security from physical layer perspectives, traditional AI approaches frequently struggle, primarily due to their limited capacity to dynamically adjust to the evolving physical attributes of transmission channels and the complexity of contemporary cyber threats. This adaptability and analytical depth are precisely where GAI excels. Therefore, in this paper, we offer an extensive survey on the various applications of GAI in enhancing security within the physical layer of communication networks. We first emphasize the importance of advanced GAI models in this area, including Generative Adversarial Networks (GANs), Autoencoders (AEs), Variational Autoencoders (VAEs), and Diffusion Models (DMs). We delve into the roles of GAI in addressing challenges of physical layer security, focusing on communication confidentiality, authentication, availability, resilience, and integrity. Furthermore, we also present future research directions focusing model improvements, multi-scenario deployment, resource-efficient optimization, and secure semantic communication, highlighting the multifaceted potential of GAI to address emerging challenges in secure physical layer communications and sensing.

\end{abstract}
\begin{IEEEkeywords}
Generative AI, physical layer communications, physical layer security, wireless sensor network, anomaly detection. 
\end{IEEEkeywords}
\IEEEpeerreviewmaketitle

\section{Introduction}\label{intro}

Generative Artificial Intelligence (GAI) represents a transformative category of Artificial Intelligence (AI) technologies capable of creating content, ranging from text, images, music, to complex simulations \cite{baidoo2023education}. As a kind of unsupervised learning, GAI is trained on vast amounts of data to understand the underlying structure and dynamics of that data. Unlike traditional AI, which primarily focuses on analyzing and interpreting data, GAI takes a step further by generating new, original outputs based on learned patterns and datasets \cite{du2023age}. Once trained, these models can produce outputs that mimic the original data's style, tone, and complexity, often indistinguishable from content produced by humans \cite{cao2023comprehensive}. Reflecting on its inherent capabilities, GAI has been successfully deployed in a wide range of mature applications across different fields, including Stable Diffusion \cite{rombach2021highresolution}, DALL-E 3 \cite{BetkerImprovingIG}, and ChatGPT \cite{wu2023brief}, etc. Beyond its prowess in generating varied content forms, GAI also demonstrates powerful capabilities in enhancing cybersecurity measures, by generating sophisticated simulations and datasets for threat detection and system strengthening \cite{dutta2020generative}. The innovative essence and wide-ranging applicability of GAI have captivated the research community, leading to an upsurge in interest to uncover its capabilities in addressing intricate challenges and driving innovation across various fields.

In wireless communications, security is a critical aspect of modern information technology, ensuring the confidentiality, integrity, and availability of data transmitted across networks \cite{aldossary2016data}. Techniques such as encryption, secure socket layers, and digital signatures are employed to protect sensitive information during its transmission over the internet or other communication networks \cite{zhang2023deep}. In the Open Systems Interconnection model of communications \cite{kumar2014osi}, physical layer security plays a pivotal role in protecting communication networks by utilizing the inherent physical characteristics of the communication channel to thwart unauthorized access and guarantee data integrity \cite{zhang2021physical}. This fundamental security layer capitalizes on the inherent unpredictability of channel properties, serving to enhance conventional encryption techniques by adding an extra layer of defense against eavesdropping and cyber-attacks. Given its critical significance, researchers have dedicated extensive efforts to conduct in-depth studies on physical layer security \cite{shiu2011physical}.

With the advancement of AI, the integration of Deep Learning (DL) methods has revolutionized communication security, offering enhanced capabilities for anomaly detection, automatic threat identification, and adaptive security measures based on real-time data analysis \cite{lv2021deep}. For instance, Convolutional Neural Networks (CNNs) are employed to design physical layer security techniques such as in the development of an intrusion detection system \cite{wang2017malware}, multi-user authentication \cite{liao2019novel}. In addition, Recurrent Neural Networks (RNNs) have found utility in various studies including automatic modulation classification \cite{hong2017automatic}, secure channel coding \cite{xiao2020designing}, and intrusion detection \cite{kim2015applying}.

However, traditional AI methods often fall short in addressing physical layer security challenges due to their inability to dynamically adapt to the continuously changing physical characteristics of transmission channels and the sophisticated nature of modern cyber threats \cite{wang2023generative}. Specifically, traditional AI models are typically trained on datasets from specific environments, limiting their effectiveness when deployed in unfamiliar conditions. Furthermore, the complexity and variability of noise patterns, signal interference, and channel conditions within the physical layer lead to difficulties in collecting sufficient labeled data for physical layer attacks. This challenge necessitates the development of sophisticated AI models capable of learning from and adapting to these ever-changing environmental factors, thereby ensuring the continuous maintenance of robust security measures \cite{o2017introduction}.

\begin{figure*}[htb]
\scriptsize
\centering
\tikzset{
    my node/.style={
        draw=gray,
        thick,
        font=\sffamily,
        drop shadow,
        minimum height=0.5cm,
    },
    my node level 1/.style={
        my node,
        inner color=blue!5,
        outer color=blue!10,
        minimum width=1cm,
        rounded corners=2,
    },
    my node level 2/.style={
        my node,
        inner color=green!5,
        outer color=green!10,
        minimum width=2cm,
        rounded corners=4,
    },
    my node level 3/.style={
        my node,
        inner color=orange!5,
        outer color=orange!10,
        minimum width=3cm,
    },
}
    \begin{forest}
        for tree={%
            my node,
            l sep+=8pt,
            grow'=east,
            edge={gray, thick},
            parent anchor=east,
            child anchor=west,
            tier/.option=level, 
            edge path={
            \noexpand\path [draw, \forestoption{edge}] (!u.parent anchor) -- +(10pt,0) |- (.child anchor)\forestoption{edge label};
            },
            if level=1{ 
                edge path={
            \noexpand\path [draw, \forestoption{edge}] (!u.south) -- +(10pt,0) |- (.child anchor)\forestoption{edge label};
            },
            }{},
            if level=1{my node level 1}{},
            if level=2{my node level 2}{},
            if level=3{my node level 3}{},
            align=center, 
            baseline, 
        }
        [Generative AI for Secure Physical Layer Communications and Sensing, rotate=90
        [Introduction, minimum width=4.0cm,[Related Surveys, minimum width=4.0cm
        [{GAI for Communication Networks \cite{karapantelakis2023generative}, \cite{van2023generative}, etc;\\AI for Secure Communication \cite{xu2020artificial}, \cite{sharma2023deep}, etc.
        }, minimum width=7.0cm, text height=0.53cm, base=midpoint
        ]
        ]
        ]
        [Background Knowledge, minimum width=4.0cm, base=midpoint,
        [Security Issues \\ in Physical Layer, minimum width=4.0cm
        [{Confidentiality, Authentication, Availability,\\ Resilience and Integrity
        }, minimum width=7.0cm, 
        ]
        ]
        [Overview of GAI, minimum width=4.0cm [AEs and VAEs \cite{zhai2018autoencoder}; GANs \cite{goodfellow2020generative}; DMs \cite{ho2020denoising}., minimum width=7.0cm]]
        ]
        [Communication Confidentiality \\and Authentication, minimum width=4.0cm, base=midpoint,
        [Secure Communication, minimum width=4.0cm, [{\textit{Encryption}: GAN \cite{smith2019communication}, SVAE \cite{lin2020variational}, etc.\\
        \textit{Anti-eavesdropping}: VQ-VAE \cite{nemati2023vq}, WGAN-GP \cite{han2022novel}, etc.}, minimum width=7.0cm, base=midpoint, text height=0.53cm]][Communication Authentication, minimum width=4.0cm, [
        {\textit{RF Fingerprinting}: GAN \cite{merchant2019securing}, Triple-GAN \cite{gong2019generative}, etc.\\
        \textit{CSI}: CGAN \cite{germain2021mobile}, etc. \textit{CIR}: VAE-PLA \cite{meng2022physical}, etc.
        }, minimum width=7.0cm , base=midpoint, text height=0.53cm]]
        ]
        [Communication Availability \\and Resilience, minimum width=4.0cm, base=midpoint,
        [Anti-jamming Strategy, minimum width=4.0cm, [{
        \textit{Jamming Recognition}: GAN \cite{erpek2018deep}, AC-VAEGAN \cite{tang2020jamming}, etc.\\
        \textit{Anti-jamming}: GAN \cite{han2021better}, ADRLDN \cite{wang2021double}, etc.
        }, minimum width=7.0cm, base=midpoint, text height=0.53cm]][Spoofing Defense, minimum width=4.0cm, [{GAN \cite{shi2019generative}, CBEGAN \cite{ma2022controllable}, SJG-GAN \cite{yang2022simple}, etc.}, 
        minimum width=7.0cm]]
        ]
        [Communication Integrity, minimum width=4.0cm, 
        [Anomaly Detection, minimum width=4.0cm, [{
        \textit{Score-based}: SAIFE \cite{rajendran2019unsupervised} , $\beta$-VAE \cite{harini2023data}, etc.\\
        \textit{Prediction-based}: E-GAN \cite{zhou2021radio}, CGAN \cite{toma2020ai}, etc.
        }, minimum width=7.0cm, base=midpoint, text height=0.53cm]
        ]
        [Data Reconstruction, minimum width=4.0cm, [{SARGAN \cite{tran2018generative}, VAE-GAN \cite{feng2022waveform}, CDDM \cite{wu2023cddm}, etc.}, minimum width=7.0cm]]
        ]
        [Future Research Direction, minimum width=4.0cm, 
        ]
        [Conclusion, minimum width=4.0cm, ]
        ]
    \end{forest}
\caption{The structure of the survey paper, where we introduce GAI methods for physical layer security through Communication Confidentiality and Authentication (Section \ref{CCA}), Communication Availability and Resilience (Section \ref{CAR}), and Communication Integrity (Section \ref{CI}).
}
\label{fig:my_tikz}
\end{figure*}
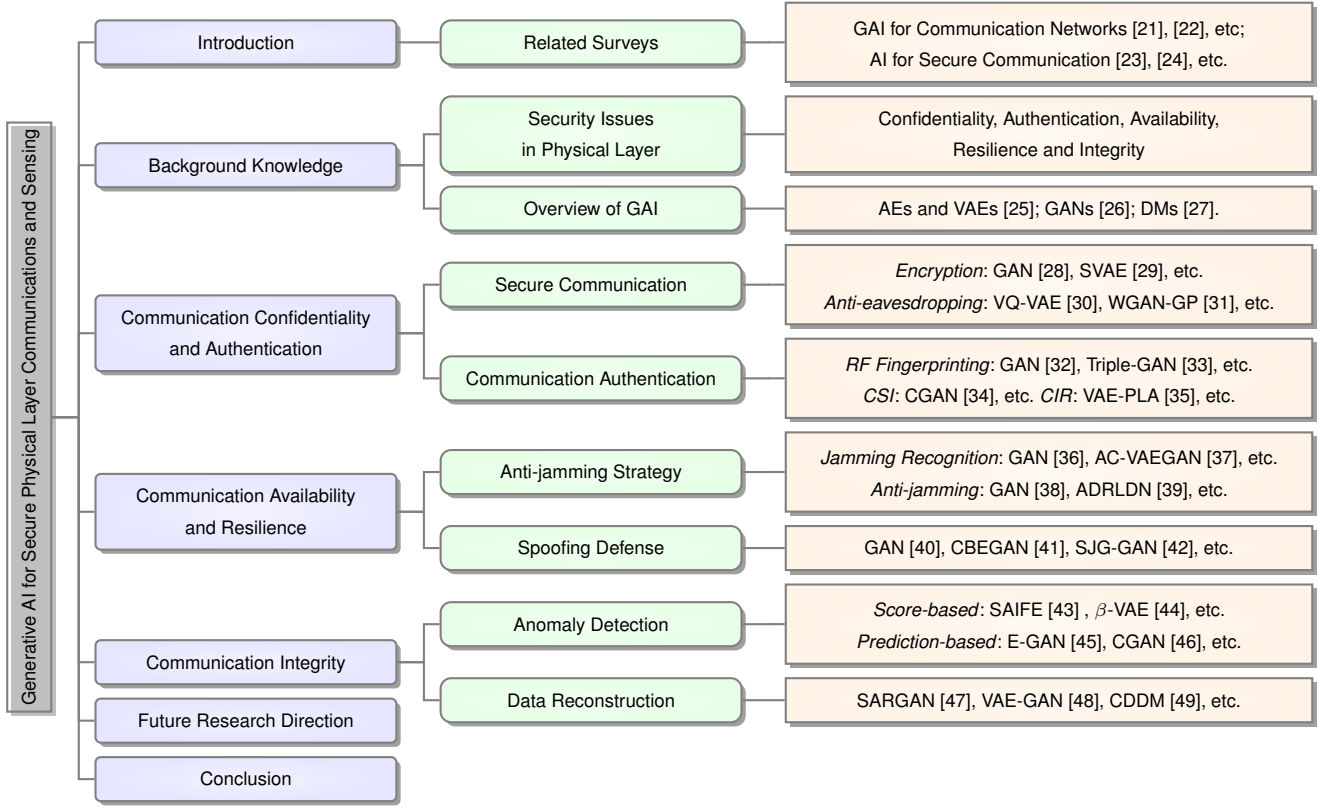

Confronted with the critical challenges in secure physical layer communications, and recognizing the distinct advantages provided by GAI, this paper provides a thorough survey of GAI's applications in tackling various issues in physical layer security. 

\subsection{Related Surveys and Contribution}

\begin{table*}[htp]
\scriptsize
\centering
\caption{SUMMARY OF RELATED SURVEYS}
\label{tab:related}
\begin{tabular}{m{0.12\textwidth}<{\centering}||m{0.08\textwidth}<{\centering}|m{0.14\textwidth}<{\centering}|m{0.55\textwidth}}
    \hline
    \textbf{Scope} & \textbf{Reference}  & \textbf{Emphasis} & \multicolumn{1}{c}{\textbf{Overview}}  \\
    \hline
   \multirow{4}{0.12\textwidth}[-5pt]{\centering GAI for Communication Networks}& \cite{karapantelakis2023generative}& GAI in mobile networks  & A survey of the recent work in the field of GAI with application to mobile telecommunications networks \\
    \cline{2-2} \cline{3-4}
    &\cite{xu2024unleashing}&  Edge-Cloud GAI  & An overview of research activities related to AIGC, GAI, and mobile edge intelligence\\
    \cline{2-2} \cline{3-4}
    &\cite{liang2023generative}& GAI-driven SemCom & A summary on the interplay between GAI and SemCom in wireless communication networks\\
    \cline{2-2} \cline{3-4}
   & \cite{van2023generative} &   GAI for Physical Layer Communications & A survey of GAI’s applications to address diverse problems in physical layer communications\\
    \hline
    \multirow{7}{0.12\textwidth}[-15pt]{\centering AI for Secure Communication}&\cite{xu2020artificial} & AI for IoT security 
    & A summary of the contribution of AI to the IoT security in Edge computing\\
    \cline{2-2} \cline{3-4}
    &\cite{nguyen2021security} & AI for Security and Privacy of 6G
    & A overview of security and privacy issues based on prospective technologies for 6G in the physical, connection, and service layers\\
    \cline{2-2} \cline{3-4}
    &\cite{santhosh2023comprehensive}& AI-based Intrusion Detection System & A survey on machine learning-based intrusion detection systems for secure communication in IoTs\\
    \cline{2-2} \cline{3-4}
    &\cite{alwahedi2024machine}& AI in IoT Security & A overview of applying machine learning for cyber threat detection in IoT environments\\
    \cline{2-2} \cline{3-4}
    &\cite{kamboj2021machine}& AI-based Physical Layer Security
    & A summary on intelligent wireless physical layer security by concentration on physical layer authentication, antenna selection, and relay node selection\\
    \cline{2-2} \cline{3-4}
    &\cite{sharma2023ai} &  AI-assisted Secure Data Transmission
    & An in-depth analysis of the role of AI in optimizing and designing the intelligent physical layer security techniques\\
    \cline{2-2} \cline{3-4}
   & \cite{sharma2023deep}  & AI-based Physical Layer Security 
   & A survey about employing DL-based physical layer security techniques for solving various security concerns in 5G and beyond networks\\
    \hline
  \end{tabular}
\end{table*}

\subsubsection{GAI for Communication Networks}

Recent literature has witnessed a notable increase in the exploration of GAI applications within communication networks (Table \ref{tab:related}). The work \cite{karapantelakis2023generative} delves into the utilization of GAI to address contemporary challenges in mobile telecommunications networks.  This article underscores the pivotal role of generative AI in the advancement of mobile network technologies, particularly in the overcoming of existing obstacles. \cite{xu2024unleashing} shifts the focus to the deployment of Artificial Intelligence Generated Content (AIGC) in mobile networks, providing comprehensive insights into Generative AI and mobile edge intelligence. Additionally, \cite{liang2023generative} investigates the interplay between GAI and Semantic Communication (SemCom) in wireless networks. Their research demonstrates the utility of GAI in the creation, transmission, and efficient management of information within these networks. Moreover, the authors in \cite{van2023generative} present an analysis of GAI applications in the physical layer, addressing various applications but the security issues are not the main focus.

\subsubsection{AI for Secure Communication}

AI has significantly transformed the landscape of communication network security and privacy. In \cite{nguyen2021security}, a systematic overview is presented on prospective technologies for 6G networks, focusing on the physical, connection, and service layers, along with lessons learned from existing security architectures. The authors in \cite{xu2020artificial} discuss the contribution of AI to Internet of Things (IoTs) security within Edge Computing (EC) environments, particularly emphasizing AI's role in augmenting security features. Regarding attack detection, \cite{santhosh2023comprehensive} provides a detailed survey on AI-based intrusion detection systems, with a focus on securing communication within the IoT. Similarly, \cite{alwahedi2024machine} delves into machine learning applications, with a specific emphasis on cyber threat detection in IoT environments. For Physical-Layer Security, the utilization of AI in optimizing and designing intelligent physical layer security techniques is thoroughly explored in \cite{sharma2023ai}. \cite{kamboj2021machine} introduces intelligent wireless physical layer security by concentrating on physical layer authentication, antenna selection, and relay node selection. In a related vein, \cite{sharma2023deep} investigates DL based physical layer security techniques, concentrating on their application in addressing various security concerns in 5G and beyond networks. However, it is noted that there is a lack of detailed analysis regarding the role of GAI in physical layer security. 

Distinct from existing surveys and tutorials, our survey distinguishes itself by specifically focusing on the integration of GAI in secure physical layer for communication networks. Unlike previous works, which either broadly address GAI applications in communication networks or delve into AI's role in network security without a concentrated emphasis on GAI, this survey offers a unique perspective by marrying the capabilities of GAI with the requirements of physical layer security. It fills a critical gap in the literature by providing an in-depth analysis of how GAI can enhance security measures, detect and mitigate threats in the physical layer that have been previously underexplored or only briefly mentioned.

The key contributions of this paper are summarized as follows:

\begin{itemize}
    \item Our comprehensive analysis reveals how to employ GAI models to enhance key security properties such as communication confidentiality, authentication, availability, resilience, and integrity. These advancements are facilitated by GAI's ability to understand complex data distributions, perform encrypted data transformation and processing, and detect cyber threats and anomalies within the network infrastructure. This summary provides essential insights for further exploration and development of GAI applications in physical layer security.
    \item We explore how GAI addresses the challenges of data sparsity and incompleteness in physical layer security, which significantly impact the efficacy of traditional AI models. GAI's contribution to data reconstruction and augmentation showcases its unparalleled ability to enhance physical layer security, surpassing the limitations of tranditional AI approaches.
    \item We outline crucial future research directions for the applications of GAI in physical layer security, including model improvements, multi-scenario deployment, resource-efficient optimization and secure semantic communication. These directions are considered from multiple perspectives, underscoring the multifaceted potential of GAI to address emerging challenges.
\end{itemize}

The structure of this survey is outlined in Fig. \ref{fig:my_tikz}.
Section \ref{GAI} introduces the fundamental concepts of GAI and offer a review of related works. \ignore{Section \ref{sec:data} conducts an investigation on Data Reconstruction and Augmentation techniques.} In section \ref{CCA}, a comprehensive exploration into Communication Confidentiality and Authentication is presented. Section \ref{CAR} discusses approaches for Communication Availability and Resilience. Section \ref{CI} introduces GAI methods for Communication Integrity. Section \ref{OI} discusses future research directions, and Section \ref{conclu} concludes the paper. Additionally, Table \ref{tab:abbr} lists the abbreviations commonly employed throughout this survey.

\begin{table*}[htp]
\scriptsize
\centering
\caption{LIST OF ABBREVIATIONS}
\label{tab:abbr}
\renewcommand{\arraystretch}{1.0} 
  \begin{tabular}{m{0.15\textwidth}<{\centering}|m{0.30\textwidth}<{\centering}||m{0.15\textwidth}<{\centering}|m{0.30\textwidth}<{\centering}}
    \hline
    \textbf{Abbreviation} & \textbf{Description}  & \textbf{Abbreviation} & \textbf{Description}  \\
    \hline\hline
    AI &  Artificial Intelligence  & GAI & Generative Artificial Intelligence \\\hline
    CNN & Convolutional Neural Network & RNN & Recurrent Neural Networks
    \\\hline
    AIGC & Artificial Intelligence Generated Content & DL & Deep Learning 
    \\\hline
    DAI & Discriminative AI  & AE  & Autoencoder   \\\hline
    VAE & Variational Autoencoder & GAN & Generative Adversarial Network
     \\\hline
    DM & Diffusion Models & DRL & Deep Reinforcement Learning
     \\\hline
    WGAN-GP & Wasserstein GAN with Gradient Penalty & CGAN&  Conditional GAN 
    \\\hline
    ACGAN & Auxiliary Classifier GAN  &AAE&  Adversarial Autoencoder \\\hline
    SNR & Signal-to-Noise Ratio & JNR & Jamming-to-Noise Ratio
    \\\hline
    PCA& Principal Component Analysis & MIMO& Multi-Input Multi-Output  
     \\\hline
    CIR & Channel Impulse Response & CSI & Channel State Information
     \\\hline
     RF& Radio Frequency & LSTM & Long Short-Term Memory 
     \\\hline
     SU & Secondary User & PU & Primary User  
     \\\hline
    SemCom & Semantic Communication & IoT & Internet of Things 
    \\\hline
    EC & Edge Computing  & EH & Energy Harvesting
    \\\hline
    JSCC & Joint Source Channel Coding & BER & Bit Error Rate 
 \\\hline
     AWGN &  Additive White Gaussian Noise & BLER &Block Error Rate 
\\\hline
    
  \end{tabular}
\end{table*}

\section{Background Knowledge}\label{GAI}

In this section, we delve into the security challenges inherent to the physical layer of communication networks, arguing that addressing security at this foundational level is of paramount importance. Furthermore, we introduce the fundamental concepts of GAI, including its architecture, classification, and basic models.

\subsection{Security Issues in Physical Layer}

\begin{table*}[htp]
\scriptsize
\centering
\caption{THE USE OF GAI IN THE PHYSICAL LAYER AND ITS POTENTIAL SUPPORT FOR SECURITY}
\label{tab:issuse}
\begin{tabular}{m{0.10\textwidth}<{\centering}||m{0.22\textwidth}<{\centering}|m{0.16\textwidth}<{\centering}|m{0.12\textwidth}<{\centering}|m{0.27\textwidth}}
    \hline
    \multirow{2}{*}{\diagbox{\textbf{Issues}}{\textbf{Model}}} & 
    \multirow{2}{*}{\textbf{GANs}} &  \multirow{2}{*}{\textbf{AEs and VAEs}} & \multirow{2}{*}{\textbf{DMs}}
    & \multicolumn{1}{c}{
    \multirow{2}{*}{\textbf{Communication \& Sensing Perspectives}}}\\
    ~&~&~&~&~ \\
    \hline
    \textbf{Confidentiality} & \begin{itemize}[leftmargin=*]
       \item[\textcolor{blue-green}{\ding{108}}] key generation
       \item[\textcolor{blue-green}{\ding{108}}]  channel response approximations
        \item[\textcolor{blue-green}{\ding{108}}] anti-eavesdropping communications
      \vspace{-1.0em}
      \end{itemize} & 
    \begin{itemize}[leftmargin=*]
       \item[\textcolor{blue-green}{\ding{108}}] wiretap code design
       \item[\textcolor{blue-green}{\ding{108}}] transceiver design
        \item[\textcolor{blue-green}{\ding{108}}] VAE-based JSCC
      \vspace{-1.0em}
      \end{itemize}& - & \multirow{3}{*}[0pt]{\makecell[l]{ Potential benefits for communication:\\
      \textcolor{green}{\ding{51}} Robustness to the changing environment \\
      \textcolor{green}{\ding{51}} Simulate noise channel effects\\
      \textcolor{green}{\ding{51}} Utilize time-varying information\\
      \textcolor{green}{\ding{51}} Extract valuable features various data\\
      }
      } \\\cline{1-4}
    \textbf{Availability} & \begin{itemize}[leftmargin=*]
      \item[\textcolor{blue-green}{\ding{108}}]  jamming recognition
       \item[\textcolor{blue-green}{\ding{108}}] anti-jamming strategy 
      \vspace{-1.0em}
      \end{itemize} & - & - & \\\cline{1-4}
    \textbf{Resilience} & \begin{itemize}[leftmargin=*]
       \item[\textcolor{blue-green}{\ding{108}}] spoofing recognition 
       \item[\textcolor{blue-green}{\ding{108}}]  spoofing defense
      \vspace{-1.0em}
      \end{itemize} & - & - & \\\hline
    \textbf{Integrity} & \begin{itemize}[leftmargin=*]
       \item[\textcolor{blue-green}{\ding{108}}] sensors anomaly detection
       \item[\textcolor{blue-green}{\ding{108}}] signals anomaly detection
       \item[\textcolor{blue-green}{\ding{108}}] radio anomaly detection
       \item[\textcolor{blue-green}{\ding{108}}] spectral information completion
       \item[\textcolor{blue-green}{\ding{108}}] electromagnetic data reconstruction
      \vspace{-1.0em}
      \end{itemize} & \begin{itemize}[leftmargin=*]
       \item[\textcolor{blue-green}{\ding{108}}] spectrum anomaly detection
       \item[\textcolor{blue-green}{\ding{108}}] sensors anomaly detection
       \item[\textcolor{blue-green}{\ding{108}}] DSSS signals reconstruction
      \vspace{-1.0em}
      \end{itemize} & \begin{itemize}[leftmargin=*]
       \item[\textcolor{blue-green}{\ding{108}}] noise elimination
      \vspace{-1.0em}
      \end{itemize} & \multirow{2}{*}[10pt]{
      \makecell[l]{
      Potential benefits for sensing:\\
      \textcolor{green}{\ding{51}} Identify abnormal sensors\\
      \textcolor{green}{\ding{51}} Not affected by data imblance \\
      \textcolor{green}{\ding{51}} Avoid complex parametric analysis of \\the signals \\
      \textcolor{green}{\ding{51}} Not require any information\\ of the missing band locations \\
      }  
      }\\\cline{1-4}
    \textbf{Authentication}  & \begin{itemize}[leftmargin=*]
       \item[\textcolor{blue-green}{\ding{108}}] RF fingerprinting authentication
       \item[\textcolor{blue-green}{\ding{108}}]  CSI authentication
      \vspace{-1.0em}
      \end{itemize} & \begin{itemize}[leftmargin=*]
       \item[\textcolor{blue-green}{\ding{108}}] CIR authentication
      \vspace{-1.0em}
      \end{itemize} & - & \\\hline
  \end{tabular}
\end{table*}


Security at the physical layer is deemed paramount compared to other layers since it provides the foundation for all subsequent security protocols \cite{wang2023acceleration}. Therefore, a breach at this foundational level will jeopardize the entire communication system. This layer is susceptible to a broad spectrum of physical threats, including eavesdropping, jamming, and spoofing, making it a critical point of vulnerability that must be robustly protected \cite{wang2023through}. By securing the physical layer, potential attacks can be preemptively thwarted, thereby preventing attackers' initial access points for further intrusions.

The subsequent discussion will introduce the CIA triad \cite{samonas2014cia}: Confidentiality, Integrity, and Availability, alongside two additional critical focuses: Resilience and Authentication in physical layer security.

\begin{itemize}
    \item \textbf{Communication Confidentiality:} Communication confidentiality in the physical layer involves the use of techniques and mechanisms to secure data transmission over communication channels, preventing unauthorized access and eavesdropping. This approach leverages the properties of the physical layer, such as noise and signal characteristics, to enhance security by making it difficult for attackers to intercept or decode the transmitted information \cite{hamamreh2018classifications}.
    \item \textbf{Communication Authentication:} Communication authentication at the physical layer is a critical security measure that verifies the identities of entities engaged in data exchange to thwart impersonation and unauthorized access. This verification leverages unique attributes intrinsic to the transmission's physical medium, such as radio-frequency fingerprints or specific channel properties. By authenticating that the communication is genuine and emanates from a verified source, this process significantly bolsters security and integrity in the exchange of data \cite{bai2020physical}.
    
    \item \textbf{Communication Availability:} 
    Ensuring communication availability at the physical layer, particularly through anti-jamming measures, involves deploying strategies and mechanisms to protect wireless communication networks from deliberate interference or jamming attacks. Techniques such as frequency hopping and direct sequence spread spectrum are pivotal, as they disperse the signal across a broader bandwidth, complicating the attacker's ability to disrupt communications \cite{shiu2011physical}. 
    
    \item \textbf{Communication Resilience:} 
    Communication resilience at the physical layer, particularly in safeguarding against a range of attacks, where spoofing attacks being a typical example, necessitates the implementation of strategies aimed at detecting and neutralizing attacks signals. Central to this defensive approach is the use of unique physical features or signatures, such as Radio Frequency (RF) fingerprints or channel state information. By exploiting these intrinsic properties, networks are equipped to distinguish authentic signals from those fabricated by spoofers, significantly maintaining secure and reliable communications \cite{shahriar2014phy}.
    
    \item \textbf{Communication Integrity:} 
    To safeguard communication integrity at the physical layer, it is essential to detect anomalous data and complete missing information \cite{shakiba2021physical}. Techniques such as DL algorithms are employed to learn normal behavior patterns and subsequently identify outliers or irregularities in real-time data flows. Furthermore, data reconstruction techniques are applied to correct or mitigate the impact of these anomalies and incomplete data, guaranteeing the precise and dependable transmission of information \cite{shen2015missing}.
 
\end{itemize}

\subsection{Overview of GAI}

GAI aims to learn the underlying features of input data to generate new content that is similar to real data, in contrast to Discriminative AI (DAI), which focuses on predicting the probability or labels of data. GAI is capable of generating a wide variety of data, including text, images, videos, and so on \cite{baidoo2023education}. Usually, these generated outputs are refereed as AIGC. With the widespread adoption of AIGC, there has been a significant boost in the efficiency of content creation, even revolutionizing the production paradigms of several companies and individual creators.

Currently, GAI models used in commnication networks can be categorized as follows:
\begin{itemize}
    \item \textbf{Autoencoder (AEs) and Variational Autoencoders (VAEs):} 
    An Autoencoder is a type of artificial neural network used to learn efficient codings of unlabeled data in an unsupervised manner \cite{zhai2018autoencoder}. It works by compressing the input into a lower-dimensional code and then reconstructing the output from this representation as close as possible to the original input. By shifting from a deterministic encoding process to a probabilistic one, VAEs can learn to represent input data as a distribution in latent space \cite{doersch2016tutorial}. Through latent distributions, VAEs generate new instances that resemble the input data by sampling from the learned distribution in the latent space, making them highly effective in tasks including image generation, data augmentation, and anomaly detection \cite{oussidi2018deep}. Building upon the foundational principles of VAEs, more advanced variants such as the Vector Quantized-Variational Autoencoder (VQ-VAE) have been developed, which incorporates a discrete latent representation through vector quantization to improve the generation quality \cite{van2017neural}. In summary, AEs and VAEs offer significant benefits, including their ability to learn complex distributions and generate new data points. However, they have limitations such as the tendency to produce blurry outputs, and challenges in balancing the reconstruction fidelity and the latent space regularization during training \cite{cao2023comprehensive}.
    
    \item \textbf{Generative Adversarial Networks (GANs):} Generative Adversarial Network model is a form of unsupervised learning \cite{goodfellow2020generative}. Within a GAN, the generator network is responsible for creating data, and concurrently, the discriminator network assesses the authenticity of this generated data. Using an adversarial mechanism, the discriminator is trained to discern between real and fake data, while the generator aims to produce data that is indiscernible from real data. GANs have evolved into diverse variants, each enhancing the original concept for specific purposes. The Conditional GAN (CGAN) introduces conditionality to direct the generative process with greater precision \cite{mirza2014conditional}. In parallel, the Wasserstein GAN with Gradient Penalty (WGAN-GP) marks a significant stride in stabilizing the training process, adeptly countering the prevalent issue of mode collapse with its innovative loss function \cite{arjovsky2017wasserstein}. The Auxiliary Classifier GAN (ACGAN) ingeniously integrates an auxiliary classifier into the discriminator, thereby elevating the fidelity and diversity of the generated images \cite{odena2017conditional}. The Adversarial Autoencoder (AAE) forges a pathway by blending AEs with adversarial training, enforcing specialized distributions within the latent space \cite{makhzani2015adversarial}. The Variational Auto-Encoding Generative Adversarial Network (VAEGAN) synergizes the structured latent spaces of VAEs with the superior generation capabilities of GANs \cite{larsen2016autoencoding}. 
    \ignore{Culminating these advancements, the BicycleGAN emerges as a hybrid, skillfully amalgamating the virtues of VAEs with GANs, particularly excelling in multimodal image-to-image translation \cite{zhu2017toward}.}
    Collectively, GANs excel in generating high-quality, realistic content and learning data distributions without explicit modeling. However, they are challenged by training difficulties, potential for mode collapse, and the generation of nonsensical outputs. 

    \item \textbf{Diffusion Models (DMs):} Diffusion Model, also known as the score-based generative model, is a novel type of generative model inspired by non-equilibrium thermodynamics \cite{ho2020denoising}. 
    Similarly to VAE, DMs aim to learn the distribution of the original data. By adding noise to the original data, the data distribution can approach a normal distribution. Through denoising steps, the noise from the normal distribution is reverted back to data from the original distribution. 
    \ignore{
    Diffusion models typically require long training times due to their iterative refinement process and the computational complexity of simulating detailed data distributions. To accelerate the sampling time of DMs, the Denoising Diffusion Implicit Models (DDIM) enable faster and more efficient generation of high-quality images through a non-Markovian process \cite{song2020denoising}. Moreover, Conditional Diffusion Models (CDMs) extend the capabilities of traditional DMs by incorporating conditioning information, such as class labels or text descriptions, allowing for guided and targeted generation of data \cite{batzolis2021conditional}. Latent Diffusion Models represent a further advancement, operating in a compressed latent space to enhance the efficiency and scalability of diffusion models, especially in high-resolution image generation tasks \cite{rombach2022high}.}
    Diffusion models are proficient in generating highly realistic data, but their long training times and computational intensity highlight the potential for further efficiency improvements in this emerging method \cite{rombach2022high}.
    
\end{itemize}

Given the powerful generative capabilities, GAI has been deployed in a multitude of applications, including image and video synthesis, data augmentation and so on. More recently, the integration of GAI into physical layer security in wireless communications is a burgeoning field with promising potential. To be more specific, GAI can play several roles in enhancing security at the physical layer: 1) Encrypted Communication; 2) Signals Authentication; 3) Attacks Defense; 4) Anomaly Detection; 5) Adaptive Signal Processing (Table \ref{tab:issuse}).





\section{Communication Confidentiality and Authentication}
\label{CCA}

In wireless communications, the principles of confidentiality and authentication stand as critical pillars ensuring the security and integrity of transmitted information \cite{shakiba2021physical}. However, cyber threats, such as eavesdropping and unauthenticated attacks in physical layer, significantly compromise communication security, leading to unauthorized access and information breaches \cite{zou2016survey}. This section provides an overview of employing GAI techniques to ensure communication confidentiality and authenticity.

\subsection{Secure Communication}

\begin{table*}[htp] \scriptsize
  \centering
  \caption{Summary of GAI for Secure Communication in Physical Layer \\Blue circles describe the methods; Green correct markers and Red cross markers represent pros and cons respectively.}
  \label{tab:secure}
    \begin{tabular}{m{0.11\textwidth}<{\centering}||m{0.06\textwidth}<{\centering}|m{0.09\textwidth}<{\centering}|m{0.64\textwidth}<{\centering}}
      \hline
      \textbf{Techniques}  &  \textbf{Reference} & \textbf{Algorithm} & \textbf{Pros \& Cons} \\
       \hline
             \multirow{3}{0.11\textwidth}[-25pt]{\centering Encrypted Communication} &\cite{besser2019flexible} & AE & \begin{itemize}[leftmargin=*] 
            \item [\textcolor{blue-green}{\ding{108}}] A flexible wiretap code design for Gaussian wiretap channels under finite blocklength by neural network AEs
          \item[\textcolor{green}{\ding{51}}] Flexibility to trade-off between the BLER and leakage.
          \item[\textcolor{green}{\ding{51}}] Achieve decent performance with simple network structures.
      \item[\textcolor{red}{\ding{55}}] Slightly worse performance than the polar wiretap codes.
      \item[\textcolor{red}{\ding{55}}] Trained with fixed SNR.
      \vspace{-1.0em}
      \end{itemize}\\
      \cline{2-4}
      &  \cite{smith2019communication} & GAN & \begin{itemize}[leftmargin=*]
    \item[\textcolor{blue-green}{\ding{108}}] The GAN architecture to learn non-linearities, memory effects, and non-Gaussian statistics.
    \item[\textcolor{green}{\ding{51}}] Approximate the channel response under different conditions.
      \item[\textcolor{red}{\ding{55}}] Based on a specific distribution dataset.
      \vspace{-1.0em}
      \end{itemize}\\
      \cline{2-4}
      & \cite{lin2020variational}  & SVAE  & \begin{itemize}[leftmargin=*]
      \item[\textcolor{blue-green}{\ding{108}}] A DL-based transceiver design for secrecy systems
          \item[\textcolor{green}{\ding{51}}] Novel loss to measure the information leakage
          \item[\textcolor{green}{\ding{51}}] Robustness to the changing environment
          \item[\textcolor{green}{\ding{51}}] Trained in an unsupervised fashion without labeling effort
      \item[\textcolor{red}{\ding{55}}] Limited scalability of the code.
      \item[\textcolor{red}{\ding{55}}] Learning high-dimensional codes is computationally challenging.
      \vspace{-1.0em}
      \end{itemize}\\
      \hline
      \multirow{4}{0.11\textwidth}[-25pt]{\centering Anti-eavesdropping} 
      &  \cite{erdemir2022privacy}  & VAE  & \begin{itemize}[leftmargin=*]
\item[\textcolor{blue-green}{\ding{108}}]
      A data-driven approach using VAE-based JSCC
          \item[\textcolor{green}{\ding{51}}] A joint source channel coding framework
          \item[\textcolor{green}{\ding{51}}] Hide the sensitive information different from the original signal
      \item[\textcolor{red}{\ding{55}}] The eavesdropper’s channel quality is assumed to be significantly worse.
      \vspace{-1.0em}
      \end{itemize}\\
      \cline{2-4}
      & \cite{nemati2023vq} & VQ-VAE & \begin{itemize}[leftmargin=*]
      \item[\textcolor{blue-green}{\ding{108}}] A JSCC scheme
based on VQ-VAE for point-to-point wireless communication
          \item[\textcolor{green}{\ding{51}}] Simulate noise channel effects
          \item[\textcolor{green}{\ding{51}}] Learn joint codewords incorporating the characteristics of the message channel
      \item[\textcolor{red}{\ding{55}}] Highly dependent on training set
      \vspace{-1.0em}
      \end{itemize}\\
      \cline{2-4}
       & \cite{fadul2021adversarial}  & GAN & \begin{itemize}[leftmargin=*]
       \item[\textcolor{blue-green}{\ding{108}}] A GAN inspired approach using DSSS to ensure that a transmitter and receiver can communicate safe 
          \item[\textcolor{green}{\ding{51}}] Use low Peak Side Lobe to improve model convergence
          \item[\textcolor{green}{\ding{51}}] Use of multiple spreading codes instead of one shared code
      \item[\textcolor{red}{\ding{55}}]  Relatively high reconstruction accuracy of eavesdroppers
      \vspace{-1.0em}
      \end{itemize}\\
      \cline{2-4}
      &  \cite{han2022novel} & WGAN-GP  & \begin{itemize}[leftmargin=*]
      \item[\textcolor{blue-green}{\ding{108}}] A physical layer key generation method based on WGAN-GP AAE
          \item[\textcolor{green}{\ding{51}}] Overcome the difficulty of quantifying the extracted features
          \item[\textcolor{green}{\ding{51}}] Reduce the quantization complexity
      \item[\textcolor{red}{\ding{55}}] The key randomness is related to the interpretability of neural network. 
      \vspace{-1.0em}
      \end{itemize}\\
      \hline
    \end{tabular}
\end{table*}

Eavesdropping is a typical attack in physical layer which involves intercepting and accessing confidential information transmitted over networks \cite{sheikh2019survey}. To improve confidentiality and achieve anti-eavesdropping, secure data is usually encrypted through various encryption algorithms \cite{shi2021physical}. However, once the math problem used for encryption is solved effective, the security of the encryption method will be seriously compromised. Moreover, several transitional methods including Error-Correcting Codes (ECCs)  \cite{sweeney1991error} suffer from a dilemma that they cannot achieve the trade-off between the reliability and data leakage because of the fixed code parameters. GAI methods, particularly those employing AEs or VAEs, offer enhanced security via generating complex structures that are difficult to decipher or reverse-engineer. 

\begin{figure}[htp]
    \centering
    \includegraphics[width= 0.95\linewidth]{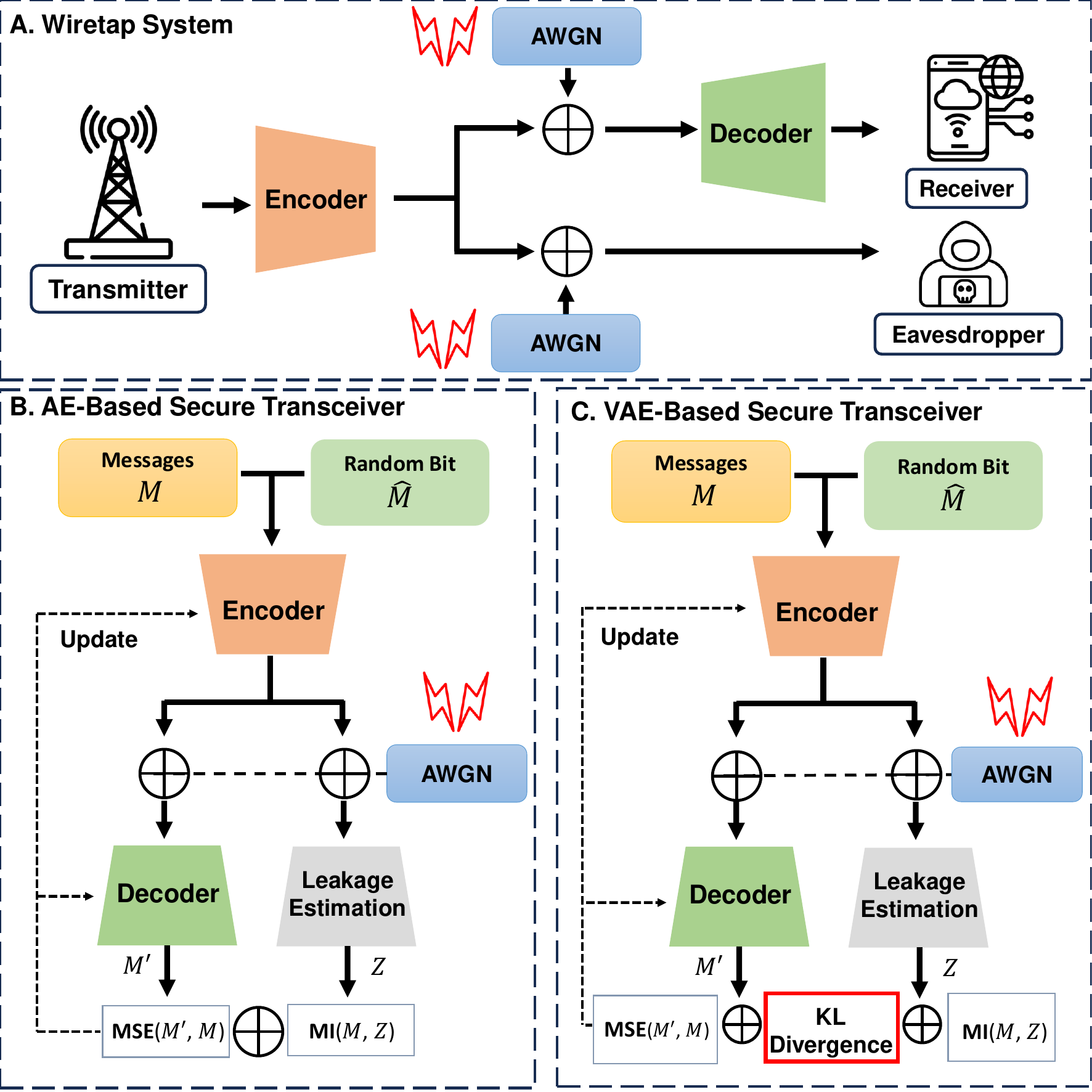}
    \caption{The overall architecture of the AE-Based \cite{besser2019flexible} and VAE-Based secure transceiver \cite{lin2020variational}.
    \textit{Part A} demonstrates a wiretap system model with AWGN. \textit{Part B} illustrates the whole framework of AE-Based secure transceiver, which is trained by two loss functions: the mean-squared error between transmitter messages and reconstructed messages and the mutual information between the messages and the received symbols by the eavesdropper. In \textit{Part C}, the VAE-Based secure transceiver adds additional loss function: KL divergence.}
    \label{fig:AE+VAE}
\end{figure}

AEs, characterized by its encoder and decoder components, enable the efficient encoding of information into a compressed, less interpretable format for transmission. Based on this, the authors proposed a AE-based framework in \cite{besser2019flexible}, which allows a flexible design of finite blocklength wiretap codes (Fig. \ref{fig:AE+VAE}). The operating point with respect to the trade-off between Block Error Rate (BLER) and information leakage can be changed easily due to the higher flexibility. In the scenario with Additive White Gaussian Noise (AWGN) channels as noise \cite{besser2019flexible}, all tested AEs perform slightly worse than the polar wiretap code \cite{mahdavifar2011achieving}. The proposed framework achieve a BLER of around 26.2\% and a leakage around 1.46bits while the polar wiretap code achieves a leakage around 1.33bits at a similar BLER of around 26.4\% \cite{besser2019flexible}. Even though the performance is not better than the polar wiretap code, the proposed framework can take the advantage of the flexibility to trade-off between the BLER and leakage, and may improve performance with deeper neural networks.

To train a secure communication system based on AEs, loss gradients need to be passed backward from the output layer of the receiver to the input layer of the transmitter. However, a practical challenge arises due to the unknown gradients of the physical channel. This issue is often circumvented by assuming channel models with known analytic expressions, such as AWGN and Rayleigh fading channels \cite{sklar1997rayleigh}. However, these models may not accurately reflect real-world channel conditions. To overcome this limitation, the authors in \cite{smith2019communication} proposed a communication channel density estimating GAN, inspired by BicycleGAN \cite{zhu2017toward}. The proposed method focuses on channels characterized by a combination of non-linear amplifier distortion, pulse shape filtering, inter-symbol interference, frequency-dependent group delay, multipath, and non-Gaussian statistics. They conducted a comparative analysis of the marginalized probability density functions of the channel with a trained generator. Through experiments conducted on four different channels, the results indicate that the proposed model is capable of generating high-accuracy approximations of the channel \cite{smith2019communication}. 

One of the shortcomings of this AE-based framework is that it is trained based on the fixed Signal-to-Noise Ratio (SNR). When the SNR varies in the testing phase, their method cannot provide the optimal solution. To address this issue, the authors in \cite{lin2020variational} proposed a VAE-based scheme for secrecy systems in changing environment (Fig. \ref{fig:AE+VAE}). In this framework, they proposed a secure VAE (SVAE) to perform as a transceiver designed with a loss function which can measure the information leakage. \ignore{The SVAE, comprising distinct encoder and decoder components as an encryption process, is trained to be robust against AWGN. }Its loss function is specifically designed to increase the difficulty for eavesdroppers to recover data, thereby meeting diverse application requirements of the transceiver. In experiments, Bit Error Rate (BER) is adopted to evaluate the communication quality. Compared with the AE-based method \cite{besser2019flexible}, \ignore{the BER at legitimate receiver of SVAE decreases faster than AE-based with SNR increasing in a perfect CSI scenario. Particularly, }the BER of SVAE achieves around $5\times10^{-6}$ when SNR is $10$ dB where the BER of AE-based achieves $10^{-2}$ in a perfect CSI scenario. Moreover, the BER at eavesdropper of SVAE does not decrease and keep at $0.5$ where AE-based's BER has marked decrease \cite{lin2020variational}.
\ignore{
In an imperfect CSI scenario, considering different Pilot-to-Noise Ratio (PNR), the BER of iCSI-SVAE is the highest and keeps at $0.55$ where AE-based and SVAE are both at $0.5$ \cite{lin2020variational}. }With this high BER at eavesdropper and low BER at legitimate, the eavesdropper cannot recover correct information and the legitimate can keep a high communication quality.

However, the model in \cite{besser2019flexible} only focuses on channel coding, where source and channel coding are performed separately, might not be as efficient in dynamic communication networks. Joint Source Channel Coding (JSCC) can dynamically adjust the coding strategy based on both the source content and the channel conditions which is more suitable for dynamic networks \cite{farsad2018deep}. The authors in \cite{erdemir2022privacy} proposed a data-driven approach using VAE-based JSCC. The proposed model aims to minimize the information leakage and emphasises hiding an underlying sensitive information. 
\ignore{
The VAE provides substantial control over the modeling of latent distributions and facilitates the practical computation of variational approximations in the cost function. This capability is especially crucial for the proposed model to utilize the mutual information between sensitive data and the noise-altered codewords that eavesdroppers might intercept, offering a tractable calculation of information security dynamics.}
The VAE enables precise control over latent distributions and practical variational approximation computation, crucial for calculating information security dynamics in the proposed model. Evaluated on colored MNIST dataset, the proposed method provides minimally distorted source transmission with maximum channel capacity \cite{erdemir2022privacy}. 

Similarly, the authors of \cite{nemati2023vq} \ignore{delved into Data-driven JSCC. They }proposed a VQ-VAE \cite{van2017neural} based JSCC wireless communication framework. This framework interprets both channel and source encoder (ENC) and decoder (DEC) as variational techniques. A notable feature of the VQ-VAE is the codebook, which facilitates the modeling of noisy channels in communication. Specifically, noise is represented by codeword through an index of binary digits to improve generalization \cite{sklar2021digital}. 
\ignore{
In VQ-VAE, it is a standard practice to introduce noise into the codewords to improve generalization beyond training data. 
}
Similarly, distortion can be injected between the ENC and DEC to emulate a noisy channel, enhancing the quality of communication. Furthermore, the ENC, DEC, and codebook are intricately dependent on the dataset and channel conditions during the training phase \cite{nemati2023vq}. Leveraging this dependency, the framework can be adapted for secure and private communication applications specifically to prevent eavesdropping. In such scenarios, an eavesdropper would face significant challenges in accurately reconstructing the intended message without prior knowledge of the specific ENC and codebook used.

Beyond directly employing VAEs for encoding transmitted information, \cite{fadul2021adversarial} investigated the application of a GAN-inspired model for covert communication through Direct-Sequence Spread Spectrum (DSSS). This model is designed to secure communications between two parties, Alice and Bob, by preventing an eavesdropper, Eve, from detecting 
\ignore{
intercepting, and exploiting their communications }in an AWGN environment. In this setup, Alice transmits a message to Bob using a spreading code from a shared codebook.
\ignore{while Eve attempts to intercept and decode the communication without access to the codebook. }The GAN-inspired model utilizes the information eavesdropped by Eve to generate the spreading code used by Alice and Bob. The system is jointly trained with a combined loss function, aiming to minimize Bob's reconstruction error while maximizing Eve's. Furthermore, spreading sequences with low Peak Side Lobe can improve model convergence. However, when Eve employs Auto-Correlation-based Detection techniques, Eve detectd the presence of the DSSS signal with an accuracy of 70\% at -6 dB SNRs or higher \cite{fadul2021adversarial}. This significant level of detection accuracy indicates that the proposed method must adopt more proactive strategies to ensure enhanced security in communications.


Key generation that exploits the unpredictable characteristics of wireless channels can provide information-theoretic security for communication confidentiality \cite{zhang2016experimental}. By utilizing the unique and unpredictable characteristics of channels, key generation methods can effectively prevent eavesdroppers from gaining access to the encrypted data. However, when directly adopting AEs or VAEs, the unpredictability of the hidden layer output and the inability to estimate high-dimensional features in advance pose challenges for applying these methods to key generation in the physical layer of communication systems. Therefore, a physical layer key generation method based on WGAN-GP \cite{arjovsky2017wasserstein} based AAE was proposed in \cite{han2022novel}. This model is designed to efficiently extract features between legitimate nodes in a way that these features align with a Gaussian distribution. Compared with the Principal Component Analysis (PCA) method \cite{li2018high}, the proposed method can yield higher security key capacity and a lower key error rate 15\% which is lower than PCA with feedback, and 10\% lower than the without feedback PCA. Additionally, the key generation ratio of this method is much higher than that achieved with PCA \cite{han2022novel}. 

As summarized in Table \ref{tab:secure}, due to its ability to learn distributions and extract features, GAI can significantly improve the security of data transmission by generating encryption algorithms preventing evolving threats. However, existing encryption methods \cite{besser2019flexible, smith2019communication, lin2020variational} mostly depend on specific dataset reducing the generality of the method. Improving the generalization ability of the GAI model while maintaining accuracy may be a research direction in the future. In addition, communication quality depends greatly on the interpretability of neural networks \cite{han2022novel}, which is still an open question.

\subsection{Communication Authentication}

\begin{table*}[htp] \scriptsize
  \centering
  \caption{Summary of GAI for Communication Authentication in Physical Layer \\Blue circles describe the methods; Green correct markers and Red cross markers represent pros and cons respectively.}
  \label{tab:auth}
    \begin{tabular}{m{0.11\textwidth}<{\centering}||m{0.06\textwidth}<{\centering}|m{0.09\textwidth}<{\centering}|m{0.64\textwidth}<{\centering}}
      \hline
      \textbf{Techniques}  &  \textbf{Reference} & \textbf{Algorithm} & \textbf{Pros \& Cons} \\
       \hline
      \multirow{3}{0.11\textwidth}[-20pt]{\centering RF Fingerprinting Authentication} & \cite{merchant2019securing}  & GAN & \begin{itemize}[leftmargin=*]
\item[\textcolor{blue-green}{\ding{108}}] A GAN framework adapted for use within the RF fingerprinting context
          \item[\textcolor{green}{\ding{51}}] Introduce a weakness in conventional AI methods
          \item[\textcolor{green}{\ding{51}}] Augment the train dataset.
      \item[\textcolor{red}{\ding{55}}] Cannot handle fast time-varying information
      \vspace{-1.0em}
      \end{itemize}\\
      \cline{2-4}
      & \cite{roy2019rfal} &  GAN & \begin{itemize}[leftmargin=*]
      \item[\textcolor{blue-green}{\ding{108}}] A framework for
building a robust system to identify rogue RF transmitters
          \item[\textcolor{green}{\ding{51}}] Exploit transmitter specific “signatures” including I/Q imbalance
      \item[\textcolor{red}{\ding{55}}] Require additional classifier
      \vspace{-1.0em}
      \end{itemize}\\
      \cline{2-4}
      &\cite{han2020radio} &  GAN & \begin{itemize}[leftmargin=*]
      \item[\textcolor{blue-green}{\ding{108}}]  A robust wireless
transmitter identification scheme using GAN
          \item[\textcolor{green}{\ding{51}}] Use a multi-classifier to both detect attackers and transmitters
      \item[\textcolor{red}{\ding{55}}] Real wireless channel effect is not reflected
      \vspace{-1.0em}
      \end{itemize}\\
      \cline{2-4}
     &  \cite{gong2019generative}  &  Triple-GAN & \begin{itemize}[leftmargin=*]
     \item[\textcolor{blue-green}{\ding{108}}] A semi-supervised specific emitter identification using GAN 
          \item[\textcolor{green}{\ding{51}}] A semi-supervised classification
      \item[\textcolor{red}{\ding{55}}] Require relatively long training time
      \vspace{-1.0em}
      \end{itemize}\\
      \hline
      \multirow{2}{0.11\textwidth}[-8pt]{\centering CSI Authentication} & \cite{germain2020physical}  & GAN & \begin{itemize}[leftmargin=*]
      \item[\textcolor{blue-green}{\ding{108}}] The use of GAN and measured MIMO communications channel information to make a decision on  Authentication
          \item[\textcolor{green}{\ding{51}}] Effective in a variety of wireless environments
          \item[\textcolor{green}{\ding{51}}] Use adversarial training for the discriminative model
      \item[\textcolor{red}{\ding{55}}] Cannot handle fast time-varying information
      \item[\textcolor{red}{\ding{55}}] Relatively limited performance at low SNR
      \vspace{-1.0em}
      \end{itemize}\\
      \cline{2-4}
      & \cite{germain2021mobile}  & CGAN & \begin{itemize}[leftmargin=*]
      \item[\textcolor{blue-green}{\ding{108}}] A method for physical layer
authentication using two variations of CGAN
          \item[\textcolor{green}{\ding{51}}] Utilize time-varying CSI as conditional input
          \item[\textcolor{green}{\ding{51}}] Handle stochastic nature of the wireless channel
      \item[\textcolor{red}{\ding{55}}] Not outstanding performance.
      \vspace{-1.0em}
      \end{itemize}\\
      \hline
      \multirow{1}{0.11\textwidth}[0pt]{\centering CIR Authentication} & \cite{meng2022physical}  &   HVAE-PLA  & \begin{itemize}[leftmargin=*]
      \item[\textcolor{blue-green}{\ding{108}}]  HVAE algorithm applied for learning industrial wireless channels
          \item[\textcolor{green}{\ding{51}}] Without requiring attackers’ prior channel information
          \item[\textcolor{green}{\ding{51}}] Extract valuable features of high-dimensional CIRs
      \item[\textcolor{red}{\ding{55}}] Require relatively long training time
      \item[\textcolor{red}{\ding{55}}] A trade-off between the involved class number and the authentication performance
      \vspace{-1.0em}
      \end{itemize}\\
      \hline
    \end{tabular}
\end{table*}

In communication networks, safety-critical messages, which are essential for the safe operation and coordination of systems, are frequently transmitted. These include vital communications such as collision warnings, speed limit notifications, and updates on traffic conditions in vehicular networks \cite{azam2021comprehensive, zhang2023generative}. To ensure these messages are genuine and trustworthy, implementing an authentication process is a critical measure to thwart malicious activities. 

\ignore{Authentication serves as a security checkpoint to verify the identity of users or systems, effectively distinguishing legitimate entities from impostors.}

RF fingerprinting is a technique used to identify and authenticate wireless devices based on the distinctive characteristics inherent in RF signals. Given its ability to accurately pinpoint the source of a transmission, RF fingerprinting is seen as a crucial tool for device authentication and access control \cite{jagannath2022comprehensive}. Recently, several traditional AI methods have been adopted as the standard approach for RF fingerprinting. In \cite{merchant2018deep}, a CNN framework was proposed to distinguish transmitters by the estimated error present in their transmitted waveforms. A Long Short-Term Memory (LSTM) network model was proposed in \cite{das2018deep}. 

However, the authors in \cite{merchant2019securing} revealed a weakness in the training processes of these approaches that a malicious GAN can be trained to introduce signal imperfections without modifying the bandwidth or data contents of the signal to force classifier errors. Then they showed that the classifier, trained by the augmented dataset with adversarial examples from GAN, can mitigate this vulnerability. The experiment results demonstrate that the Receiver Operating Characteristic (ROC) curves with GAN-augmented training has nearly 1 Area Under the Curve (AUC), where 90\% of the networks without GAN perform even worse than random guessing at 40 dB SNR \cite{merchant2019securing}. Similarly, in \cite{roy2019rfal} the Radio Frequency Adversarial Learning (RFAL) framework was proposed for building a robust system to identify rogue RF transmitters by designing and implementing a GAN. The GAN utilizes the In-phase and Quadrature (IQ) imbalance \cite{jagannath2022comprehensive} to extract unique high-dimensional features from the RF signals\ignore{ (Fig. \ref{fig:RF})}. Using the augmented data from the generator in GAN, a discriminator model can classify the trusted transmitters and the counterfeit ones with 99.9\% accuracy. 

Inspired by \cite{roy2019rfal}, the authors in \cite{han2020radio} proposed a GAN based wireless transmitter identification scheme. The proposed framework uses a multi-classifier to both detect malicious attacker and classify trusted transmitter without any extra classifier. 
\ignore{
Within the GAN framework, each discriminator is trained as a binary classifier. }Once trained, discriminators are employed to check whether the captured unknown IQ data comes from a corresponding trusted transmitter. If the label vector, made up of 1s and 0s, shows all 0s, the data is not from a trusted source, suggesting a high probability of it being sourced from an attacker. Additionally, the authors in \cite{gong2019generative} introduced the Triple-GAN structure \cite{gan2017triangle} to adopt semi-supervised classification. With the modified structure, the classification accuracy of the proposed framework can achieve over 90\%, only 1\% of the training data samples are labeled \cite{gong2019generative}.



Channel State Information (CSI) is an important parameter of a communication link in the physical layer. By leveraging its unique properties, CSI can be utilized in an authentication context \cite{wang2021survey, zhang2023generative}. Once a transmitter is initially authenticated by certain methods such as RF fingerprinting, the receiver maintains this authentication status as long as the variations in the received CSI remain below a certain threshold compared to the CSI from previous transmissions. However, attackers can modify various aspects of their transmission setup, including antenna properties, transmission timing, power levels, or use reflectors \cite{siegman1966antenna}. These alterations enable them to change their CSI as measured by the receiver. 

To address these issues, the author in \cite{germain2020physical} proposed a GAN based model to authenticate devices in Multi-Input Multi-Output (MIMO) communication systems. The proposed model employs adversarial training to improve the authentication process. The discriminative model at the receiver is trained by a generator that creates fake CSI samples looking like the authentic samples. Simulation results show that the discriminator achieves 100\% accuracy for SNR greater than or equal to 10 dB. For SNR less than 10 dB, while the discriminator makes errors in correctly recognizing legitimate samples, it consistently succeeds in preventing illegitimate samples from being authenticated \cite{germain2020physical}. 

\ignore{In vehicular networks, specifically in Vehicular Ad-hoc Networks (VANETs) \cite{manvi2017survey}, CSI becomes variable due to the fast-changing nature of the environment. This dynamic environment poses significant challenges for authentication processes. }

\begin{figure}[htp]
    \centering
    \includegraphics[width= 0.95\linewidth]{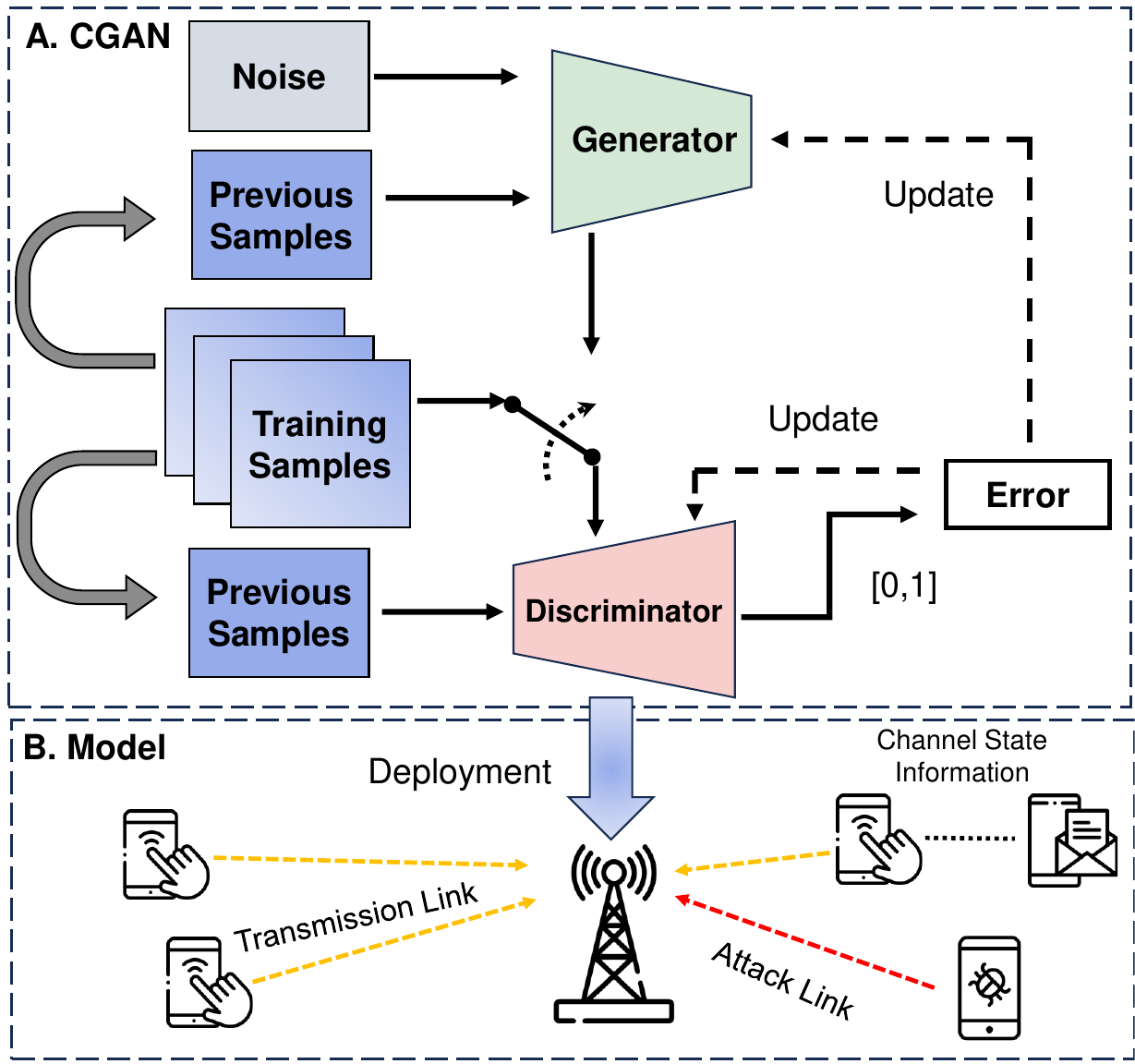}
    \caption{Proposed CGAN training architecture in \cite{germain2021mobile}. 
    In \textit{Part A}, the conditional information is the previous magnitudes of the CSI elements associated with time. The output of the discriminator is the probability value, representing the likelihood from zero to one based on its perception of whether the sample is fake or authentic. \textit{Part B} illustrates the system model structure.
    }
    \label{fig:CSI}
\end{figure}

To handle the time-varying CSI in fast-changing environment, the authors in \cite{germain2021mobile} proposed a CGAN based model combining with LSTM and gated recurrent unit (GRU) cells. Compared with the method in \cite{germain2020physical}, the proposed model utilizes a CGAN instead of a conventional one, which can incorporate the previous CSI elements associated with time as the conditional information (Fig. \ref{fig:CSI}). This approach allows for a more detailed generation and analysis of CSI data in the temporal aspect and historical patterns. In experximents, the CGAN-GRU network typically performed as well as or better than the standalone LSTM or GRU networks. Especially when mean-square error threshold is -25 dB, all of them can achieve accuracy at almost 99\% \cite{germain2021mobile}.

In addition to CSI, the Channel Impulse Response (CIR) is another significant parameter in wireless communications providing a detailed characterization of how a wireless signal propagates from the transmitter to the receiver in a specific environment. In \cite{meng2022physical}, the authors proposed a CIR-based hierarchical VAE physical-layer authentication (HVAE-PLA) scheme. The HVAE-PLA consists of an AE module and a VAE module. The AE module is dedicated to extracting the characteristics of CIR, providing insights into how signals propagate in specific environments. The VAE module building upon this aims to enhance the representational capacity of the extracted CIR characteristics. 
\ignore{Additionally, a new loss function is designed for the VAE module focusing on both the Simple Gaussian Prior and the double peak Gaussian Prior distributions for further security and robustness enhancement \cite{meng2022physical}. Finally, the results of the authentication process is outputted by the VAE module, leveraging the enhanced understanding of CIR.} Compared with a conventional AI method in \cite{xia2021multiple}, the proposed scheme can authenticate the spoofing nodes in all positions in the static dataset. Moreover, the simulations show that the proposed method can improve the authentication performance by 17.18\%–69.3\% compared to the vanilla AEs and VAEs \cite{meng2022physical}.

The integration of GAI in communication authentication through specific information has highlighted GAI's ability to enhance the uniqueness and reliability of identifying devices in a network, as summarized in Table \ref{tab:auth}. However, generative models require relatively long training and inference times due to their complex structure  \cite{gong2019generative}. When facing fast time-varying information \cite{merchant2019securing}, they have difficulty to infer and adapt to additional new information in real time. Therefore, pruning the model size and enhancing model real-time adaptation are urgently needed for security authorization.

\section{Communication Availability and Resilience}\label{CAR}

The concepts of communication availability and resilience emerge as fundamental components to maintain continuous and reliable access for communication systems \cite{bishop2011resilience}. Challenges such as network disruptions and deliberate cyber attacks can severely impact the availability of digital communication services, leading to significant downtime and loss of connectivity \cite{wang2011survey}. This section aims to explore the integration of advanced strategies and GAI techniques to ensure the communication availability and resilience via solving two common cyber attacks: jamming and spoofing in physical layers.

\subsection{Anti-jamming Strategy}

\begin{table*}[htp] \scriptsize
  \centering
  \caption{Summary of GAI for Anti-jamming Strategy in Physical Layer \\Blue circles describe the methods; Green correct markers and Red cross markers represent pros and cons respectively.}
  \label{tab:jamming}
    \begin{tabular}{m{0.11\textwidth}<{\centering}||m{0.06\textwidth}<{\centering}|m{0.09\textwidth}<{\centering}|m{0.64\textwidth}<{\centering}}
      \hline
      \textbf{Techniques}  &  \textbf{Reference} & \textbf{Algorithm} & \textbf{Pros \& Cons} \\
       \hline
      \multirow{4}{0.11\textwidth}[-30pt]{\centering Jamming Recognition} &
      \cite{erpek2018deep}  & GAN & \begin{itemize}[leftmargin=*]
      \item[\textcolor{blue-green}{\ding{108}}] An adversarial machine learning approach launch jamming attacks on wireless communications and a defense strategy
          \item[\textcolor{green}{\ding{51}}] Rely on a small amount of sample data
          \item[\textcolor{green}{\ding{51}}] Do not need any knowledge of transmitter’s algorithm
      \item[\textcolor{red}{\ding{55}}] Limited resilience of the strategy in highly dynamic and unpredictable wireless environments
      \vspace{-1.0em}
      \end{itemize}\\
      \cline{2-4}
      & \cite{de2022multi}  & GAN & \begin{itemize}[leftmargin=*]
      \item[\textcolor{blue-green}{\ding{108}}] An input-agnostic adversarial attack technique based on GANs and multi-task loss
          \item[\textcolor{green}{\ding{51}}]  Quickly craft small imperceptible perturbations
          \item[\textcolor{green}{\ding{51}}]  Not depend on the original samples
      \item[\textcolor{red}{\ding{55}}] Just consider certain scenarios
      \vspace{-1.0em}
      \end{itemize}\\
      \cline{2-4}
     &  \cite{tang2020jamming} & AC-VAEGAN & \begin{itemize}[leftmargin=*]
     \item[\textcolor{blue-green}{\ding{108}}] A jamming recognition method based on AC-VAEGAN 
          \item[\textcolor{green}{\ding{51}}] Stable recognition performance in the case of small samples
      \item[\textcolor{red}{\ding{55}}] Require relatively long training time
      \item[\textcolor{red}{\ding{55}}] Unsuitable for large-scale deployment
      \vspace{-1.0em} 
      \end{itemize}\\
      \cline{2-4}
             & \cite{cai2020spectrum}  & GAN & \begin{itemize}[leftmargin=*]
             \item[\textcolor{blue-green}{\ding{108}}] An algorithm based on GAN TO mine the relationship of the data and complete the missing data
          \item[\textcolor{green}{\ding{51}}]   Complete spectrum data in multiple jamming patterns
      \item[\textcolor{red}{\ding{55}}] Less emphasis on real-world testing
      \vspace{-1.0em}
      \end{itemize}\\
      \hline
        \multirow{6}{0.11\textwidth}[-35pt]{\centering Anti-Jamming} &
      \cite{han2021better} & GAN & \begin{itemize}[leftmargin=*]
      \item[\textcolor{blue-green}{\ding{108}}] A GAN based spectrum completion network
          \item[\textcolor{green}{\ding{51}}] Complete the partially missing spectrum data
          \item[\textcolor{green}{\ding{51}}] Achieve high reward of the policy
      \item[\textcolor{red}{\ding{55}}] Slightly poor performance in the high missing rate scenario
      \vspace{-1.0em}
      \end{itemize}\\
      \cline{2-4}
      & \cite{han2021primary} & GAN & \begin{itemize}[leftmargin=*]
      \item[\textcolor{blue-green}{\ding{108}}] A PU-friendly dynamic
spectrum anti-jamming access scheme combining offline training and online deployment
          \item[\textcolor{green}{\ding{51}}] Focus on both PU and SU
          \item[\textcolor{green}{\ding{51}}] Converge fast to the optimal policy
      \item[\textcolor{red}{\ding{55}}] Still slow convergence speed
      \vspace{-1.0em}
      \end{itemize}\\
      \cline{2-4}
      & \cite{wang2021double}  & ADRLDN & \begin{itemize}[leftmargin=*]
      \item[\textcolor{blue-green}{\ding{108}}] A decision related judgment module between jammer and user based on GAN
          \item[\textcolor{green}{\ding{51}}] Adapt to complex types of jamming
          \item[\textcolor{green}{\ding{51}}] Superior in anti-jamming performance than the current anti-jamming method
      \item[\textcolor{red}{\ding{55}}] Require relatively long training time
      \vspace{-1.0em}
      \end{itemize}\\
      \cline{2-4}
        & \cite{lin2023physical} & GAN & \begin{itemize}[leftmargin=*]
        \item[\textcolor{blue-green}{\ding{108}}] The secrecy communication in an EH-enabled Cognitive EH-CIoT network with a cooperative jammer
          \item[\textcolor{green}{\ding{51}}]  Maximize the system's secrecy rate while minimizing the SOP
      \item[\textcolor{red}{\ding{55}}] Potential limitations in real-world implementation
        \item[\textcolor{red}{\ding{55}}] complexity of the DRL framework
      \vspace{-1.0em}
      \end{itemize}\\
      \cline{2-4}
      & \cite{tang2023gan} & GAN & \begin{itemize}[leftmargin=*]
      \item[\textcolor{blue-green}{\ding{108}}] An intelligent jamming and anti-jamming framework to analyze and promote the security of semantic communication
          \item[\textcolor{green}{\ding{51}}]  A GAN-like game strategy to reflect the relationship between the semantic jammer and receiver
      \item[\textcolor{red}{\ding{55}}] Not suitable for other multilingual model
      \vspace{-1.0em}
      \end{itemize}\\
      \cline{2-4}
      & \cite{jayabalan2023generative} & GDNN & \begin{itemize}[leftmargin=*]
      \item[\textcolor{blue-green}{\ding{108}}] A communication model in cognitive radios using machine learning to learn the dynamics of jamming attacks
          \item[\textcolor{green}{\ding{51}}]  Adapt to the dynamics of the spectrum
      \item[\textcolor{red}{\ding{55}}] Require relatively long training time
      \vspace{-1.0em}
      \end{itemize}\\
      \hline
    \end{tabular}
\end{table*}



The jamming attack is a vital threat to communication availability at the physical layer, aimed at disrupting legitimate communications by introducing noise \cite{huo2017jamming}. Therefore, to ensure communication availability, detecting and mitigating jamming attacks represents a critical initial defense. There are several conventional methods employed wireless jamming attacks, including random and sensing-based jamming \cite{pirayesh2022jamming}. However, with the increasing integration of machine learning techniques into communication systems, both legitimate transmitters and malicious jammers leverage machine learning algorithms to understand the spectrum environment better which introduces new emerging types of attacks including adversarial attacks. 

\ignore{
Adversarial attacks introduce subtly crafted perturbations that are imperceptible to human observers but can significantly disrupt the functioning of systems. Specifically, these attacks target deep learning-based modulation classifiers on receivers in networks that they lead to incorrect classification results \cite{erpek2018deep, de2022multi}. 
}

In \cite{erpek2018deep}, the authors present an adversarial learning strategy employing GAN to facilitate adversarial jamming attacks. This approach enables jammers to generate synthetic data based on a small number of real data samples. These synthetic samples are then integrated into the training dataset. Simulation results indicate that the detection accuracy of a jammer closely approximates, within 0.19\% for misdetection and 3.14\% for false alarms, that of a jammer trained with a larger dataset of real samples collected over a long duration \cite{erpek2018deep}. Furthermore, based on the attack characteristics, they proposed a defense strategy for the transmitter, centered on rendering its behavior unpredictable. This can be achieved by the transmitter intentionally performing incorrect actions, such as transmitting on a busy channel or refraining from transmitting on an idle channel, during strategically selected time slots. Additionally, the authors in \cite{de2022multi} proposed an input-agnostic adversarial attack technique, which adopts GAN to create perturbations in advance. These pre-generated perturbations can then be efficiently applied to a variety of incoming signals. Furthermore, this approach has the potential to substantially aid in the development of classifiers that exhibit robustness against adversarial jamming attacks.

\ignore{While traditional AI can extract many separable features \cite{wu2017jamming},}

In the case of small sample datasets, the performance of autonomous feature extraction and classification of DL will be reduced \cite{wu2017jamming}. Especially in real-world network communications, it is difficult to obtain enough sample data for anti-jamming due to the privacy policy and the inadequacy of technical methods. To generate more realistic data, the authors in \cite{tang2020jamming} proposed a jamming recognition method based on AC-VAEGAN, which combines the VAEGAN \cite{larsen2016autoencoding} and ACGAN \cite{odena2017conditional}. In this model, the latent space of a small amount of signal dataset is obtained by VAE firstly. Then, the datasets will be expanded by sampling points of the latent space and decoding them. Finally, the discriminator of the GAN framework is extracted for jamming recognition. In experiments, when the Jamming-to-Noise Ratio (JNR) is -10dB, the average correct recognition rate of AC-VAEGAN network is approximately 65\%, where the rate of ACGAN and CNN network is only about 55\% \cite{tang2020jamming}. 

\ignore{
The higher correct recognition rate demonstrates that AC-VAEGAN can ensure the quality of generated data than ACGAN which does not consist of the encoder and decoder. 
}

\begin{figure}[htp]
    \centering
    \includegraphics[width= 0.95\linewidth]{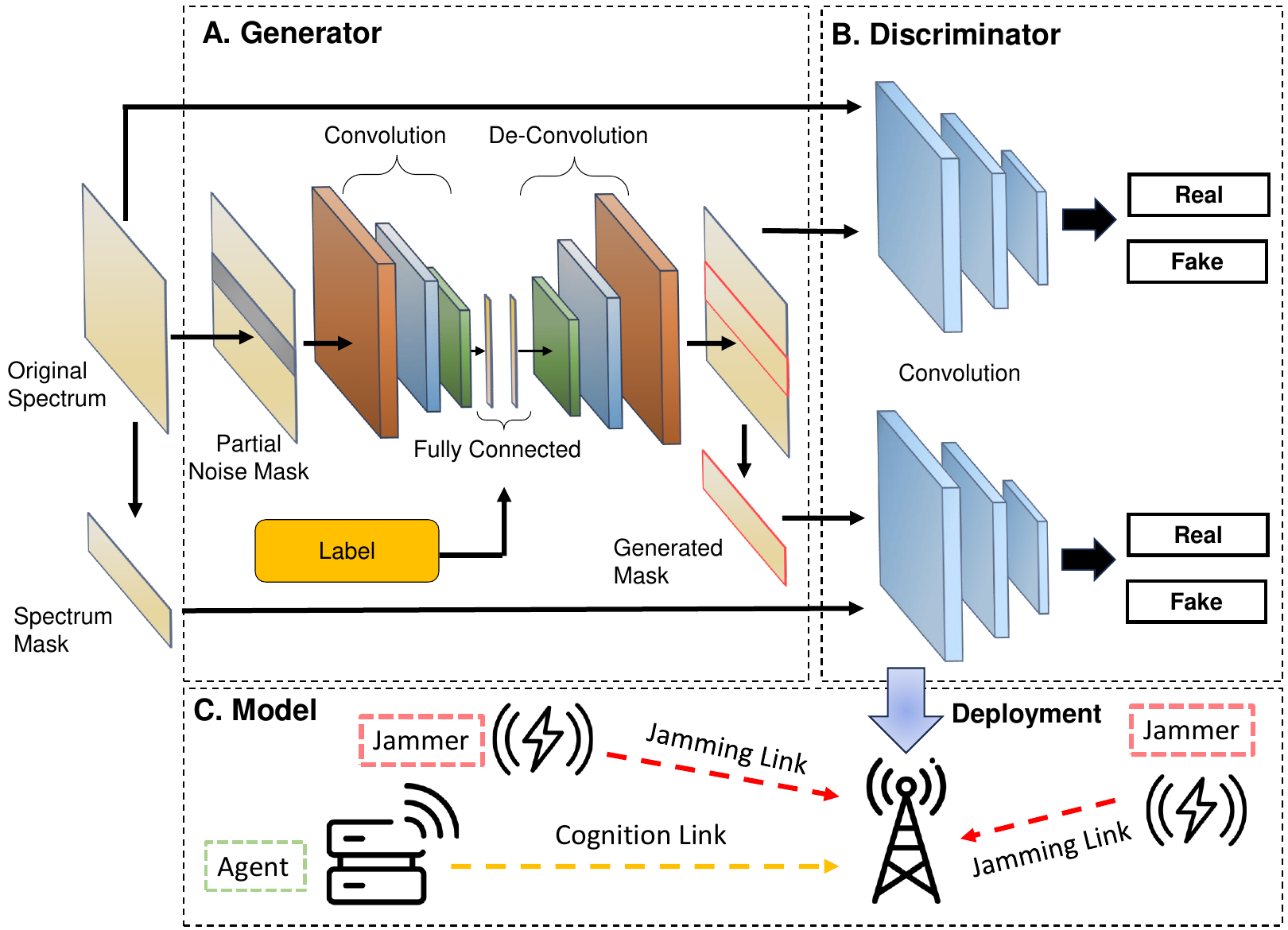}
    \caption{ The overall network structure in \cite{cai2020spectrum}.
    \textit{Part A} illustrates the generator, which is designed as an AE comprising a convolution layer, a fully connected layer, and a de-convolution layer. In \textit{Part B}, two discriminator modules are crafted using a convolution network and are optimized to focus on local and global details, respectively. \textit{Part B} describes the system model structure.}
    \label{fig:SWC}
\end{figure}

\ignore{
Although detection algorithms are typically developed to function with complete spectrum information, the occurrence of jamming attacks frequently leads to incomplete data.
}
Except the scarcity of data, the occurrence of jamming attacks frequently leads to incomplete data, which hinders the ability of anti-jamming strategies to discern jamming attacks \cite{erpek2018deep}.
\ignore{
The partial absence of spectrum data hinders the ability of anti-jamming strategies to discern and learn the regular patterns of jamming attacks \cite{erpek2018deep}.}
To complete the missing information, the authors in \cite{cai2020spectrum} proposed an efficient algorithm based on a GAN focusing on spectrum waterfall completion, where spectrum waterfall is a thermodynamic block diagram defining the environmental state \cite{cai2019jamming}. The algorithm can automatically mine the relationship of the data and complete the missing data accurately. Different from the noise input in the original GAN, they use the spectrum waterfall with missing data as generator input, which can limit generator artistry (Fig. \ref{fig:SWC}). From the corresponding complement results, the proposed algorithm is better than the method without pre-classification, since the generator that adds auxiliary information is more targeted to the data. The accuracy is more than 95\%, where the latter is nearly 80\% \cite{cai2020spectrum}.

\begin{figure*}[htp]
    \centering
    \includegraphics[width= 0.75\linewidth]{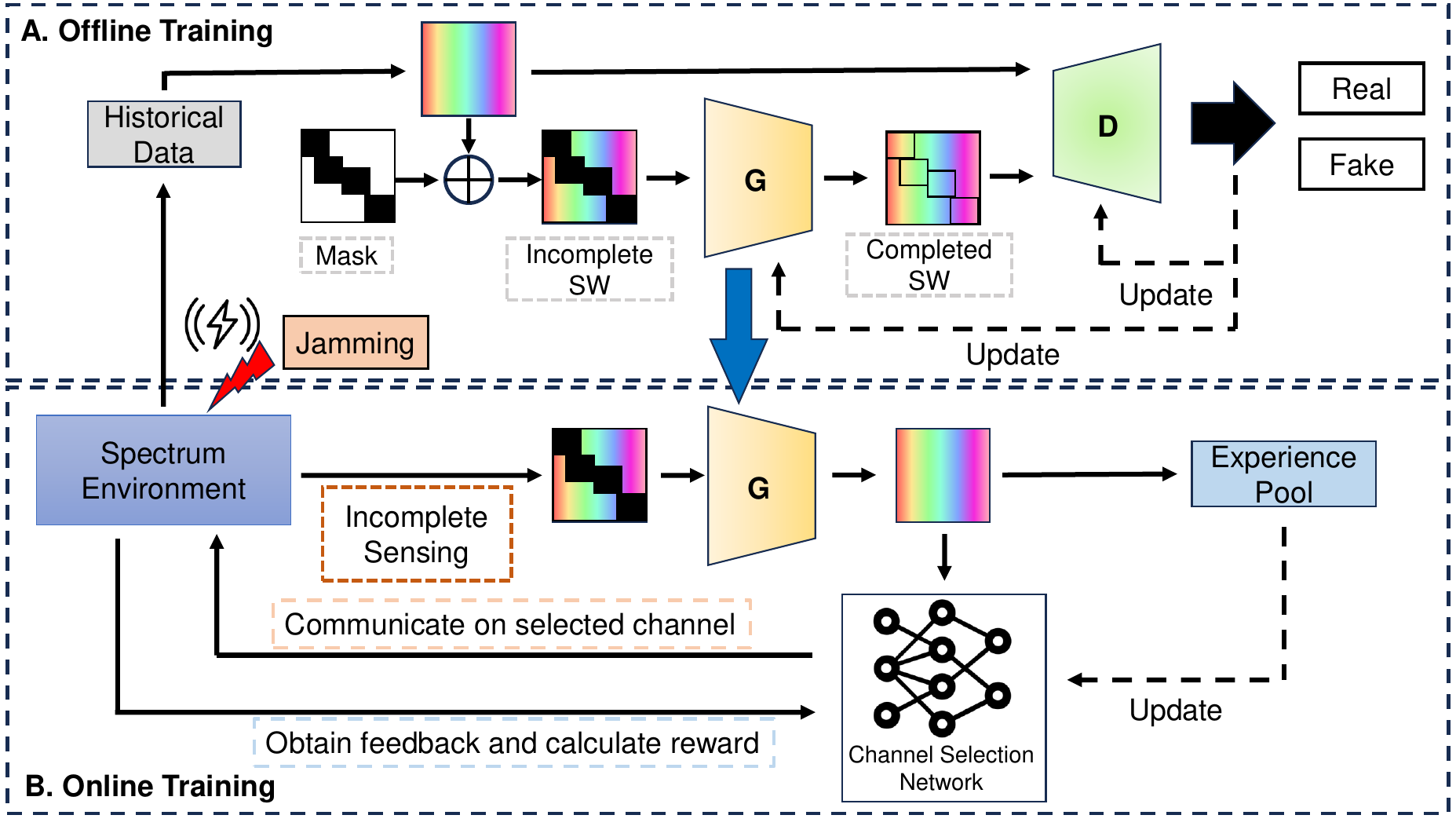}
    \caption{Overall structure of the proposed anti-jamming spectrum access scheme in \cite{han2021better}.
    In \textit{Part A}, a GAN with the Generator (G) and Discriminator (D) is trained to complete missing spectrum data using historical data. In \textit{Part B}, the generator (G) is implemented as SCN. The CSN utilizes the enriched spectrum data to assist users in selecting optimal communication channels for anti-jamming.}
    \label{fig:RL}
\end{figure*}

Several studies have demonstrated that anti-jamming communications, enhanced by Deep Reinforcement Learning (DRL) \cite{arulkumaran2017deep}, can achieve near-optimal performance in dynamic and unpredictable environments \cite{wang2020dynamic, liu2018anti}. 
\ignore{
It is important to note that the majority of these studies assume the availability of perfect spectrum information to users. However, in the dynamic environments of vehicular networks, incomplete sensing is a common occurrence, often resulting from hardware failures or transmission losses. 
In \cite{han2021better}, the authors proposed a framework designed to address and remedy this problem. }
In \cite{han2021better}, the authors proposed a framework combining the GAN and DRL. 
\ignore{
In \cite{han2021better}, a GAN-based spectrum completion network (SCN) was proposed to complete the missing spectrum data. }
The proposed framework consists of two stages: the offline stage and online stage (Fig. \ref{fig:RL}). Initially, a GAN is trained to complete missing spectrum data using historical data. Once the GAN training is complete, the generator is implemented as Spectrum Completion Network (SCN) during the online stage. Subsequently, with the augmented spectrum data, a DRL-based Channel Selection Network (CSN) is employed. The CSN utilizes the enriched spectrum data to assist users in selecting optimal communication channels for anti-jamming. The performance of the proposed scheme notably surpasses that of the conventional DRL-based method in \cite{liu2018anti}, as well as the scheme that combines K-Nearest-Neighbor Interpolation (KNNI) and DRL \cite{li2019sparsely} in all missing rates. Especially, the proposed scheme achieves the discounted accumulative reward of 8.3 when the missing rate is 10\%. In comparison, the conventional method scores 4.4, and the KNNI-DRL combination scores 6.5 \cite{han2021better}. 

Additionally, the authors in \cite{han2021primary} introduced a dynamic spectrum anti-jamming access scheme in the cognitive radio based network \cite{kavaiya2020physical} that is friendly to Primary Users (PU) while also safeguarding Secondary Users (SU) from indiscriminate jamming attack by jammers. Similar to the scheme in \cite{han2021better}, this proposed framework is divided into two stages: the offline stage and the online stage where the key difference lies in the GAN model used in the first stage. Here, the GAN is trained to accurately simulate the Spectrum Environment (SE), which is considered a Virtual Environment (VE). By pre-training the Channel Decision Network (CDN) offline in this VE, the SU is equipped to evade both PU signals and jamming in the actual SE, following the guidance of the trained CDN. According to the experiments, it takes about 90s for the proposed scheme to converge to the optimal policy while the CDN trained in SE from scratch spends about 160s \cite{han2021primary}. 
\ignore{
This arrangement, that pre-trains the CDN offline in GAN-based VE, significantly accelerates the training process and allows the CDN to converge more rapidly to the optimal policy. The VE thus acts as a sophisticated simulation platform, enabling the SU to learn and adapt to various scenarios without the risk of jammers.
}

Existing anti-jamming technologies rely on hidden anti-jamming strategies, but
\ignore{which can effectively prevent user information leakage when countering jamming attacks. However, }their performance tends to diminish when facing with more sophisticated or complex jamming types \cite{wang2020hidden}. In \cite{wang2021double}, the authors proposed a DRL algorithm with a double network structure, named ADRLDN, which adopts the hidden anti-jamming idea and can deal with various types of complex jamming in actual scenarios. In this framework, they designed a GAN network-based user and jammer decision-making correlation judgment module. The GAN is trained to fit the environmental state under known user information, and evaluates whether the user information is obtained by the jammer. The DRL network is trained to guarantee the user's decision not obtained by jammers. In this situation, there are two key points: compare the fitted environmental state with the real environmental state; ensure both the generation and the evaluation of the effect of the network. These happen to be the essential characteristics of GAN networks. According to the simulation experiment results, ADRLDN is superior in anti-jamming performance than the current anti-jamming method based on avoiding the idea (ADRLA) \cite{liu2018anti} by reducing the probability of users being jammed by 15\%. 

As for Cognitive Internet of Things (CIoT), a major challenge is extending the system's lifespan. Energy Harvesting (EH) technology is a promising solution to provide sustainable energy to energy-constrained mobile devices in CIoT systems \cite{gurjar2018wireless,xu2019secure}. However, EH-CIoT systems encounter significant jamming attacks due to the wireless channels, which exposes information transmissions to potential security risks. In \cite{lin2023physical}, the study considered an EH-CIoT system where the communication security of the SU network is threatened. To enhance security, the authors propose a DRL algorithm that integrates LSTM and GAN models. This algorithm aims to maximize the system's secrecy rate while minimizing the Secrecy Outage Probability (SOP). The GAN network is utilized to mitigate the time-varying CSI and the adverse effects of random noise at the receivers. Simultaneously, the LSTM network is employed for extracting features from the input environment. The study's findings reveal that the convergence speed of the proposed algorithm is significantly faster 1.69 and 3.15 times than that of other algorithms that do not incorporate the GAN model \cite{lin2023physical}. The integration of the GAN and LSTM model significantly enhances the algorithm's ability to quickly capture environmental information and learn optimal strategies.



SemCom is a revolutionary way of communicating that helps overcome the limitations of previous methods by using DL to send necessary information, which reduces the amount of data sent \cite{du2023semantic}. With the focus on semantic-level transmission, new challenges in jamming and anti-jamming arise. Attackers will aim to create more effective jamming methods to degrade the quality-of-experience (QoE) for users in communications \cite{du2023rethinking}. In \cite{tang2023gan}, a framework for intelligent jamming and anti-jamming in semantic communication was proposed based on the GAN. In the framework, the transmitter sends data with semantic features, and the receiver tries to understand it correctly while a jammer tries to mess up this process. The authors designed a GAN model where the jammer learns to generate disruptive signals, the receiver is trained to selectively focus on legitimate segments of the incoming data, thereby enhancing its proficiency in identifying and mitigating semantic jamming. 
\ignore{
According to the experiment results, the improved receiver has a solid anti-jamming ability through the proposed approaches, and the proposed strategy makes this semantic communication system more robust.
}

In jamming attacks, a key factor for their success is the jammer's ability to accurately determine the frequency of signal transmission. This capability is vital for generating jamming noise powerful enough to disrupt the SNR within the same frequency band. To mitigate jamming attacks, the author in \cite{jayabalan2023generative} developed an anti-jamming communication system model based on GANs. The proposed model employs generator and discriminator integrated with min-max game theory, to automatically adapt to the dynamics of the spectrum. The training of the proposed model within this defense mechanism is designed to mislead jammers, preventing them from effectively targeting the transmission of data. This strategic deception, rooted in game theory, hinders the jammers' ability to accurately select time slots for their attacks, leading to erroneous predictions on classification sources, which in turn prevents significant transmission losses. 
\ignore{
As a result, the jammers make poor decisions, which in turn prevents significant transmission losses. 
}

As summarized in Table \ref{tab:jamming}, GAI has demonstrated the effectiveness in identifying jamming activities and developing suitable anti-jamming strategies. However, most existing works only consider certain scenarios and lack real-world testing \cite{de2022multi, cai2020spectrum, tang2023gan}. Due to the complex architectures, GAI models are also unsuitable for large-scale deployment \cite{tang2020jamming}. Therefore, designing a model that can be used in realistic anti-jamming scenarios especially on a large scale needs to be considered.


\subsection{Spoofing Defense}

\begin{table*}[htp] \scriptsize
  \centering
  \caption{Summary of GAI for Spoofing Defense in Physical Layer \\Blue circles describe the methods; Green correct markers and Red cross markers represent pros and cons respectively.}
  \label{tab:spoof}
    \begin{tabular}{m{0.11\textwidth}<{\centering}||m{0.06\textwidth}<{\centering}|m{0.09\textwidth}<{\centering}|m{0.64\textwidth}<{\centering}}
      \hline
      \textbf{Techniques}  &  \textbf{Reference} & \textbf{Algorithm} & \textbf{Pros \& Cons} \\
       \hline
      \multirow{6}{0.11\textwidth}[-30pt]{\centering Spoofing Defense} & \cite{shi2019generative}  & GAN & \begin{itemize}[leftmargin=*]
      \item[\textcolor{blue-green}{\ding{108}}]  An approach of spoofing wireless signals by using a GAN
          \item[\textcolor{green}{\ding{51}}] Provide GAN-based model defense mechanism
      \item[\textcolor{red}{\ding{55}}] Limited to simulated environments
      \vspace{-1.0em}
      \end{itemize}\\
      \cline{2-4}
      & \cite{shi2020generative} & GAN & \begin{itemize}[leftmargin=*]
      \item[\textcolor{blue-green}{\ding{108}}]  A DL-based spoofing attack to generate synthetic wireless signals 
          \item[\textcolor{green}{\ding{51}}] Detailed analysis and implementation
      \item[\textcolor{red}{\ding{55}}] Not fully explore the potential countermeasures against such attacks
      \vspace{-1.0em}
      \end{itemize}\\
      \cline{2-4}
      &\cite{roy2019generative}  &  WGAN-GP & \begin{itemize}[leftmargin=*]
     \item[\textcolor{blue-green}{\ding{108}}]  The task of full-band spectral generation in addition to single signal generation 
          \item[\textcolor{green}{\ding{51}}] Treat LTE signals as 2D images
      \item[\textcolor{red}{\ding{55}}] Poor performance of generated signals
      \vspace{-1.0em}
      \end{itemize}\\
      \cline{2-4}
       & \cite{ma2022controllable}  & CBEGAN & \begin{itemize}[leftmargin=*]
       \item[\textcolor{blue-green}{\ding{108}}] A wireless spoofing attack scheme against the defense mechanism with adversarial DL
          \item[\textcolor{green}{\ding{51}}]  Compensate for transmission channel effects via auxiliary channel sensing
      \item[\textcolor{red}{\ding{55}}] Consider specific channel conditions
      \item[\textcolor{red}{\ding{55}}] Limited to simulated environments
      \vspace{-1.0em}
      \end{itemize}\\
      \cline{2-4}
      & \cite{li2021gnss}& GAN & \begin{itemize}[leftmargin=*]
      \item[\textcolor{blue-green}{\ding{108}}] A GNSS anti-spoofing method based on the idea of confrontation evolution of a GAN
          \item[\textcolor{green}{\ding{51}}] Detect small delay spoofing signals
          \item[\textcolor{green}{\ding{51}}] Extract features of slight differences
      \item[\textcolor{red}{\ding{55}}] The overall performance is not remarkable.
      \vspace{-1.0em}
      \end{itemize}\\
      \cline{2-4}
      &  \cite{yang2022simple}  &  SJG-GAN & \begin{itemize}[leftmargin=*]
      \item[\textcolor{blue-green}{\ding{108}}]  A generation method for spoofing jamming signals
          \item[\textcolor{green}{\ding{51}}] Learn the latent distribution of DSSS signals
          \item[\textcolor{green}{\ding{51}}] Propose a improved Pearson correlation coefficient
      \item[\textcolor{red}{\ding{55}}] Lack real-world testing
      \vspace{-1.0em}
      \end{itemize}\\
      \hline
    \end{tabular}
\end{table*}

Authenticating wireless signals at the physical layer is essential for ensuring communication resilience. Despite employing numerous features discussed in Section \ref{CCA}, wireless signal spoofing remains a pervasive threat. In spoofing, attackers insert fake identification information into genuine communications to join or corrupt the systems \cite{yilmaz2015survey}. Therefore, it enable unauthorized access and data manipulation at the physical layer, causing substantial harm to the communication resilience.

Currently, DL methods have proven effective against simple spoofing attacks. For example, the study in \cite{shi2019generative} examines a basic spoofing technique like the replay attack, which partially replicates original signals. 
\ignore{Such replay attacks are generally ineffective against deep learning-based classifiers. }
However, adversarial spoofing attacks, as discussed in \cite{erpek2018deep}, pose a more profound threat by evading traditional security measures. The research in \cite{shi2019generative} explored this issue from both the attacker's and defender's perspectives, proposing a GAN model to create indistinguishable signals. The GAN is trained to emulate the pattern of intended transmissions which significantly improves the possibility of a successful attack compared to random signal and replay attacks, even when node locations vary between training and testing phases \cite{shi2019generative}. Moreover, the proposed GAN-based model provides defense mechanism by using GAI to distinguish and counteract signal spoofing attacks. In \cite{shi2020generative}, the authors further provided a detailed analysis of the proposed GAN-based spoofing attack, including its implementation on embedded platforms. This implementation is carried out on two distinct embedded platforms: an embedded Graphics Processing Unit (GPU) \cite{nvidia_jetson_nano} and a Field-Programmable Gate Array (FPGA) \cite{xilinx_zynq_ultrascale}. The effectiveness of the proposed attack is noteworthy, with a success probability ranging from 60.6\% to 97.8\%. However, the technique's dependence on real-time compensation for transmission channels causes considerable overhead. This feature elevates the risk of detection due to an expanded communication footprint. 

\ignore{
However, the proposed attack scheme exhibits limitations, particularly its capacity to mimic only single emitters using a singular modulation type. This constraint diminishes its effectiveness in multifaceted, real-world scenarios. Furthermore, the technique's dependence on real-time compensation for transmission channels causes considerable overhead. This, in turn, elevates the risk of detection due to an expanded communication footprint. 
}

Simulating and imitating RF signals is a basic tactic employed by spoofers. While GAI has demonstrated effectiveness in augmenting short time series segments, challenges remain in accurately generating RF signals, such as the length of signals, and radio environments \cite{roy2019generative}. The authors in \cite{roy2019generative} explored the potential of GAN models to accomplish full-band spectral generation for anti-spoofing attacks. They implement the WGAN-GP model \cite{arjovsky2017wasserstein} to improve training stability. Drawing on its proven effectiveness in the image domain, they utilize spectral representations of OFDM signals called LTE \cite{ndujiuba2015comparative}, treating them as 2D images. Nonetheless, the study shows that using GANs to create long sequences over time is quite challenging. It is harder to capture small details and features in these long sequences than in the shorter ones \cite{roy2019generative}. 

To overcome the limitations of existing methods, \cite{ma2022controllable} introduces a controllable wireless spoofing attack scheme that leverages a Conditional Boundary Equilibrium Generative Adversarial Network (CBEGAN) \cite{marzouk2019conditional} in conjunction with auxiliary channel sensing. The CBEGAN network combines with an AAE, which is a well-established deep neural network architecture for computer vision-related tasks enabling learning with few samples \cite{makhzani2015adversarial}. It facilitates more precise and effective spoofing attacks by simulating a variety of emitters and modulation types. Additionally, the integration of auxiliary channel sensing effectively compensates for transmission channel effects. Since it allows the attack model to be trained offline, it significantly reduces the likelihood of detection by legitimate communication pairs. Under the same channel conditions, the proposed spoofing attack scheme reaches a success probability of 85.7\%. In contrast, the comparative attack scheme mentioned in \cite{shi2020generative} achieves a lower success rate, with a probability of 76.2\% \cite{ma2022controllable}.

\begin{figure}[htp]
    \centering
    \includegraphics[width= 0.95\linewidth]{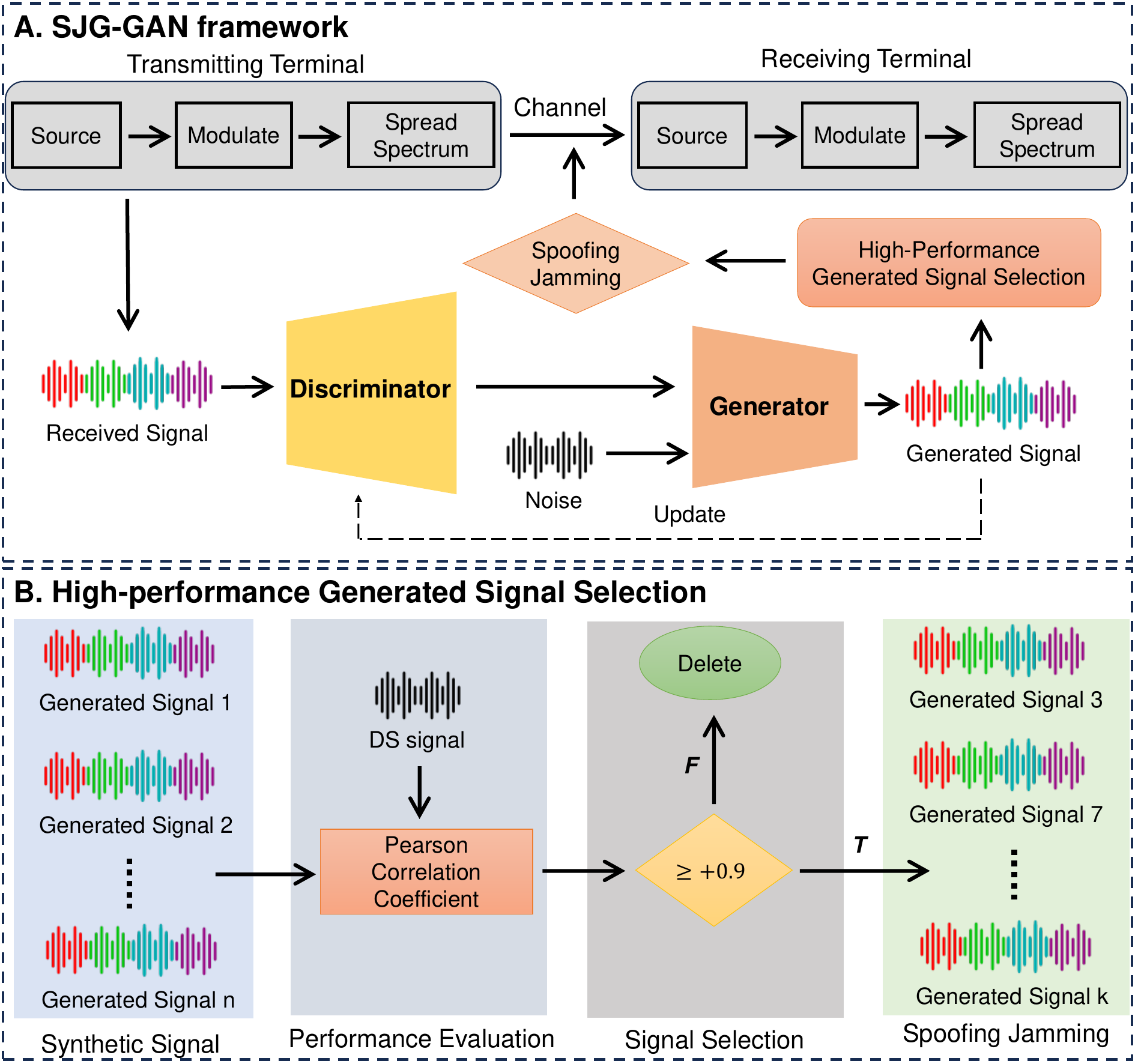}
    \caption{
    The overall network structure in \cite{yang2022simple}. \textit{Part A} illustrates the SJG-GAN framework consisting of two parts: signal generation and the high-performance generated signal selection. In \textit{Part B}, the process of the high-performance generated signal selection is shown. It can be split into two steps: the evaluation and the selection}
    \label{fig:spoof}
\end{figure}

\ignore{
One sophisticated form of spoofing attack is spoofing jamming, where the attacker broadcasts analog signals designed to imitate authentic signals. This can lead to a target receiver obtaining false information, manipulated by the attacker, instead of the true data. With the advancement of the Global Navigation Satellite System (GNSS), vehicles within networks now rely heavily on it for navigation, positioning, and timing \cite{kaplan2017understanding}. Due to their long-distance transmission from satellites to receivers, the GNSS signals are more prone to disruptions from spoofing jamming attacks. However, detecting spoofing jamming becomes particularly challenging when the spoofing signal closely resembles the authentic signal. 
}

One sophisticated form of spoofing attack is spoofing jamming, where the attacker broadcasts analog signals designed to imitate authentic signals. This can lead to a target receiver obtaining false information instead of the true data. Due to their long-distance transmission from satellites to receivers, Global Navigation Satellite System (GNSS) signals are more prone to disruptions from spoofing jamming attacks \cite{kaplan2017understanding}. To detect spoofing jamming attacks, the authors in \cite{li2021gnss} proposed a spoofing signal detection method based on the GAN in the acquisition stage, which is one of several phases in character recognition that also includes preprocessing, feature extraction, classification, and post-processing \cite{alginahi2010preprocessing}. The proposed model specifically considers the classification of authentic and spoofing signals within the context of navigation tasks. In this setup, both the training and test datasets are derived from the GPS receiver code. According to the simulation results, the successful detection of small-delay spoofing signals is achieved through the use of adversarial learning within the GAN. Additionally, while the overall performance of the GAN is comparable to that of the CNN, the GAN exhibits a slight advantage over the CNN, particularly when the pseudo-code phase offset is equal to or greater than 0.5 chip \cite{li2021gnss}. 

\ignore{
A key challenge is that the spoofing signal often exhibits only slight differences from the authentic signal. This subtle discrepancy can significantly reduce the model's performance if the discriminator is not augmented which can be done by introducing the generators in the GAN. }

However, Spoofing jamming's creation is complex and resource-intensive, requiring extensive prior information. So far, only specific authorized or civilian systems have successfully executed such attacks \cite{bhatti2017hostile}. Drawing inspiration from \cite{shi2019generative}, the authors in \cite{yang2022simple} introduced a GAN-based approach, termed Spoofing Jamming Generation (SJG-GAN), for crafting spoofing jamming attacks (Fig. \ref{fig:spoof}). This model is adept at learning the latent distribution of DSSS signals and generating a set of synthetic signals. Upon completion of training, a improved Pearson correlation coefficient is used as an evaluation metric to select the most aggressive synthetic signals for the DSSS system as spoofing jamming signals. Notably, the one-dimensional GAN model of SJG-GAN simplifies the generation process, making it more cost-effective and feasible for a variety of communication systems, as demonstrated in simulations \cite{shi2019generative}. 

In conclusions, GAI models are able to offer sophisticated mechanisms to both detect and counteract spoofing attacks, as summarized in Table \ref{tab:spoof}. However, since they are limited to simulated environments, the detection accuracy in actual scenarios still needs further investigation.

\section{Communication Integrity}\label{CI}

Communication integrity in the physical layer of a network involves ensuring that the data transmitted over a physical medium, such as copper wires, fiber-optic cables, or wireless signals, is delivered accurately and reliably, without corruption or alteration \cite{shakiba2021physical}. This often requires mechanisms for anomaly detection or data reconstruction to maintain the fidelity of the data as it moves from one device to another, thereby preserving the integrity of communication.

\subsection{Anomaly Detection}

\begin{table*}[htp] \scriptsize
  \centering
  \caption{Summary of GAI for Anomaly Detection in Physical Layer \\Blue circles describe the methods; Green correct markers and Red cross markers represent pros and cons respectively.}
  \label{tab:anomaly}
    \begin{tabular}{m{0.11\textwidth}<{\centering}||m{0.08\textwidth}<{\centering}|m{0.13\textwidth}<{\centering}|m{0.57\textwidth}<{\centering}}
      \hline
      \textbf{Techniques}  &  \textbf{Reference} & \textbf{Algorithm} & \textbf{Pros \& Cons} \\
       \hline
       \multirow{6}{0.11\textwidth}[-30pt]{\centering Score-based Detection}
       & \cite{feng2017anomaly}  & AE & \begin{itemize}[leftmargin=*]
       \item[\textcolor{blue-green}{\ding{108}}] The deep-structure AE neural networks to detect the anomalies of spectrum via time–frequency diagram
          \item[\textcolor{green}{\ding{51}}]  The threshold selected to trade-off a balance between the probabilities of false alarms and missed alarms.
      \item[\textcolor{red}{\ding{55}}] Include potential biases in signal data selection
      \vspace{-1.0em}
      \end{itemize}\\
       \cline{2-4}
     &  \cite{rajendran2019unsupervised}  & SAIFE & \begin{itemize}[leftmargin=*]
     \item[\textcolor{blue-green}{\ding{108}}] An AAE-based anomaly detector for wireless spectrum anomaly detection using PSD data
          \item[\textcolor{green}{\ding{51}}] A single model across multiple bands to extract interpretable features
      \item[\textcolor{red}{\ding{55}}] The distribution corresponds more to the latent representations than to the original training samples.
      \vspace{-2.0em}
      \end{itemize}\\
      \cline{2-4}
      & \cite{gkelias2022gan} & GAN & \begin{itemize}[leftmargin=*]
      \item[\textcolor{blue-green}{\ding{108}}]
      A GAN-based system trained on available EM signals to detect unseen types of EM waveforms
          \item[\textcolor{green}{\ding{51}}] The generator consists of two AEs connected in series.
      \item[\textcolor{red}{\ding{55}}] Require relatively long training time
      \vspace{-1.0em}
      \end{itemize}\\
      \cline{2-4}
             & \cite{cheng2022resnet} & ResNet-AE & \begin{itemize}[leftmargin=*]
             \item[\textcolor{blue-green}{\ding{108}}] An anomaly detection method based on ResNet-AE
          \item[\textcolor{green}{\ding{51}}] Establish an adaptive decision threshold
      \item[\textcolor{red}{\ding{55}}] Cannot classify the types of anomalies
      \vspace{-1.0em}
      \end{itemize}\\
       \cline{2-4}
       & \cite{harini2023data}  & $\beta$-VAE & \begin{itemize}[leftmargin=*]
       \item[\textcolor{blue-green}{\ding{108}}] 
A VAE model uses multivariate normal distribution with a parameter $\beta$ included to the KL divergence term
          \item[\textcolor{green}{\ding{51}}] Investigate the impact of different weightings of the KL divergence
      \item[\textcolor{red}{\ding{55}}] Not specify how to select the optimal value of the $\beta$ coefficient
      \vspace{-1.0em}
      \end{itemize}\\
       \cline{2-4}
      & \cite{luo2018distributed}   & AE & \begin{itemize}[leftmargin=*]
      \item[\textcolor{blue-green}{\ding{108}}] The AE neural networks into WSN to solve the anomaly detection problem
         \item[\textcolor{green}{\ding{51}}] Satisfy the demand for limited computational resources
      \item[\textcolor{red}{\ding{55}}] Lack of analysis that extends to large scale sensor networks
      \vspace{-1.0em}
      \end{itemize}\\
             \hline
      \multirow{5}{0.11\textwidth}[-30pt]{\centering Prediction-based Detection}
      &\cite{lu2021gan} & MSGAN & \begin{itemize}[leftmargin=*]
      \item[\textcolor{blue-green}{\ding{108}}] A domain-specific framework consisting of offline training and online inference to detect anomalies in the scenario of industrial robotic sensors
      \item[\textcolor{green}{\ding{51}}]  Use an adaptive update strategy during offline training
      \item[\textcolor{red}{\ding{55}}] Lack of analysis that extends to diversity scenarios
      \vspace{-1.0em}
      \end{itemize}\\
      \cline{2-4}
       & \cite{zhou2021radio} & E-GAN & \begin{itemize}[leftmargin=*]
       \item[\textcolor{blue-green}{\ding{108}}] A radio anomaly detection algorithm based on modified GAN
          \item[\textcolor{green}{\ding{51}}] Latent representations are controlled rather than being randomly selected.
          \item[\textcolor{green}{\ding{51}}] Capture the distribution of input samples
      \item[\textcolor{red}{\ding{55}}] Relatively high time complexity
      \vspace{-1.0em}
      \end{itemize}\\
       \cline{2-4}
       & \cite{toma2020ai}  &  CGAN & \begin{itemize}[leftmargin=*]
       \item[\textcolor{blue-green}{\ding{108}}] The GAI-based abnormality Detection techniques at the physical layer in CR
          \item[\textcolor{green}{\ding{51}}] Implement a hybrid structure for low- and high-dimensionality data
      \item[\textcolor{red}{\ding{55}}] Limited by signal types tested
      \vspace{-1.0em}
      \end{itemize}\\
       \cline{2-4}
        & \cite{toma2020deep} & Multiple & \begin{itemize}[leftmargin=*]
        \item[\textcolor{blue-green}{\ding{108}}]
        The GAI frameworks used to detect anomalies inside the dynamic radio spectrum
          \item[\textcolor{green}{\ding{51}}] A comparative analysis of three deep generative models
      \item[\textcolor{red}{\ding{55}}] Cannot be employed to characterize and classify the anomalous signals
      \vspace{-1.0em}
      \end{itemize}\\
      \cline{2-4}
       & \cite{rathinavel2022efficient} & Multiple & \begin{itemize}[leftmargin=*]
       \item[\textcolor{blue-green}{\ding{108}}] GAI-based  anomaly detection methods to detect a set of anomalous activities in several radio band
         \item[\textcolor{green}{\ding{51}}] Three deep generative models are applied to spectral density functions
      \item[\textcolor{red}{\ding{55}}] Need to consider the key applications and proper methods or ensembles of methods to achieve the best performance
      \vspace{-1.0em}
      \end{itemize}\\
      \hline
    \end{tabular}
\end{table*}

\begin{figure}[htp]
    \centering
    \includegraphics[width= 1.0\linewidth]{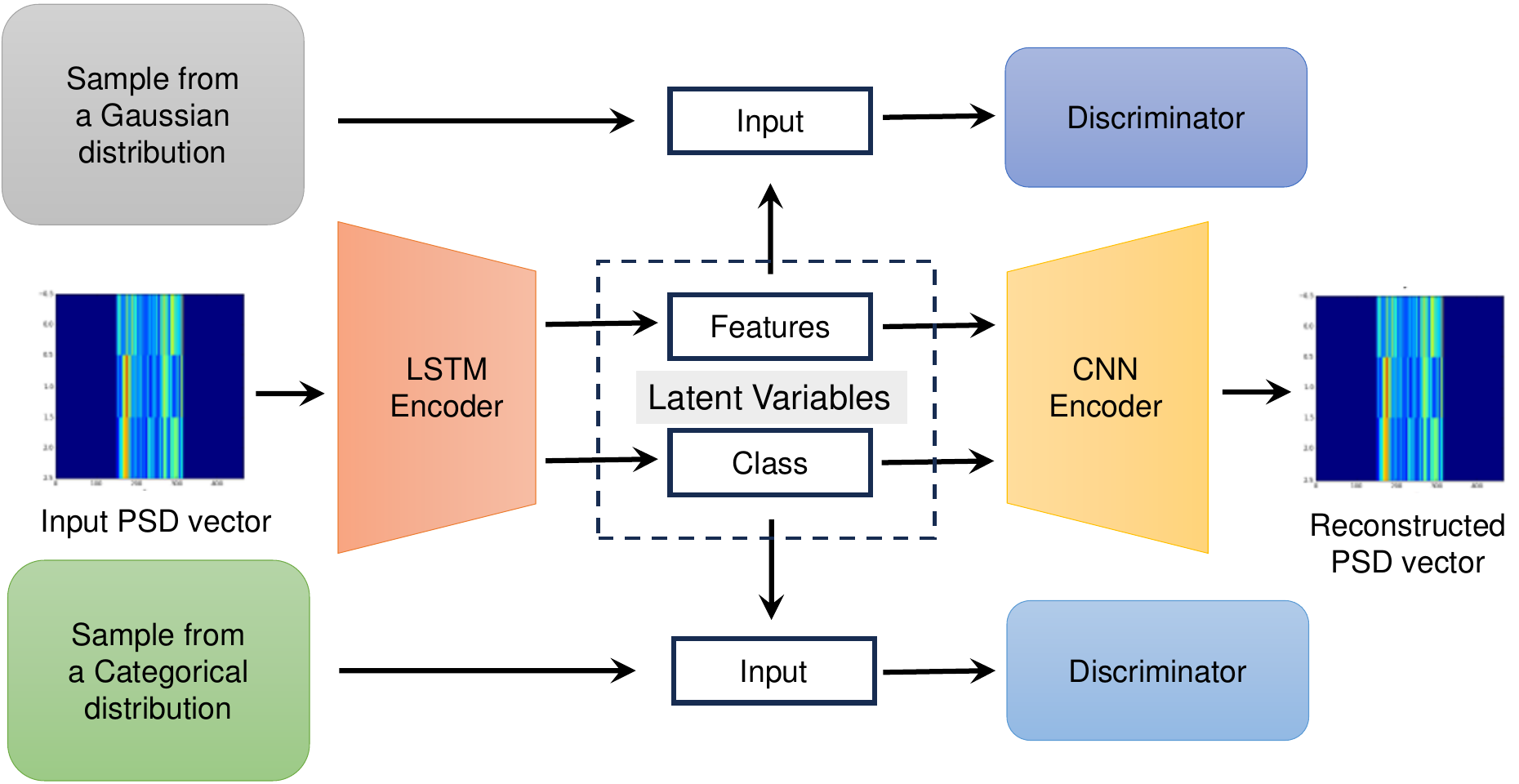}
    \caption{The proposed model architecture in \cite{rajendran2019unsupervised}. The AAE architecture is trained in a semi-supervised learning for making the features more interpretable while the reconstruction is fully unsupervised.}
    \label{fig:anomal}
\end{figure}


Anomaly detection is a method used to identify and mitigate unexpected deviations or irregularities in the communication channel \cite{chandola2009anomaly}. However, due to the unpredictable characteristics of potential anomalies, it is difficult to gather enough abnormal data for training samples in traditional AI methods \cite{el2020machine}. Thus, there is an urgent need for an unsupervised, automatic feature-extraction learning model, such as GAI techniques for anomaly detection \cite{liu2023deep}. Generally, GAI for anomaly detection can be divided into two cartography: reconstruction-based \cite{feng2017anomaly} and prediction-base detection \cite{zhou2021radio}. 

Reconstruction-based methods identify anomalies through anomaly scores, which are usually the reconstruction error in GAI models. In the study \cite{feng2017anomaly}, the authors proposed a deep AE-based approach for anomaly detection in the spectrum. The time-frequency features of preprocessed signal data are utilized to train the proposed network. To differentiate between normal and anomalous data, the method applies a threshold to the reconstruction errors, transforming these errors into a binary outcome. The threshold value is strategically selected to trade-off a balance between the probabilities of false alarms and missed alarms, and it is determined as the median of a sequence of reconstruction errors in this study. 

\ignore{
The performance of the proposed AE network is evaluated in comparison to two conventional methods: Linear Principal Component Analysis (LPCA) \cite{schein2003generalized} and Robust Principal Component Analysis (RPCA) \cite{candes2011robust}. The results demonstrate that the proposed auto-encoders model consistently surpasses these traditional approaches \cite{feng2017anomaly}.  
}
\ignore{
Notably, even in high SNR scenarios at an SNR of 25 dB, the two-layer and three-layer AE networks achieve accuracies of 71.02\% and 71.19\% respectively, where the LPCA and RPCA methods yield lower accuracies of 64.19\% and 61.91\% \cite{feng2017anomaly}. Consequently, reconstruction-based anomaly detection could be superior to prediction-based techniques, particularly in the context of wireless spectrum.
}

While the model presented in \cite{feng2017anomaly} demonstrates effectiveness, it lacks interpretable feature extraction capabilities, such as signal bandwidth and position. This limitation necessitates the training of multiple copies of the model for different frequency bands. Addressing this issues, the authors in \cite{rajendran2019unsupervised} introduced Spectrum Anomaly Detector with Interpretable FEatures (SAIFE), which is an AAE based model (Fig. \ref{fig:anomal}). SAIFE enables the training of a single model across multiple bands in an unsupervised manner, thereby eliminating the need for multiple model instances for different bands. Moreover, the AAE architecture provides a flexible and robust platform for semi-supervised learning, enabling the extraction of interpretable features based on Power Spectral Density (PSD) data \cite{elson1995calculation}. Furthermore, the reconstructed signals are a key asset for localizing anomalies within the wireless spectrum. Impressively, the model demonstrates exceptional performance in wireless band classification, achieving an accuracy close to 100\% while only utilizing 20\% labeled samples \cite{rajendran2019unsupervised}.

Similar to the SAIFE model \cite{rajendran2019unsupervised}, the study in \cite{gkelias2022gan} also designed an anomaly score based on latent representations. The authors proposed an architecture for electromagnetic waveform anomaly detection, utilizing a dual AE enhanced GAN. This design differs from the SAIFE model, as the generator in the proposed method is composed of two AEs connected in series. These encoders map original and reconstructed data to the latent space, respectively. The Anomaly Score is defined with the objective of minimizing the L2 distance between latent representations for anomaly detection. 

\ignore{
According to the simulations conducted, the proposed system has been effectively trained and tested on a synthetic dataset. The results demonstrate that the system is capable of achieving Area Under the Receiver Operating Characteristic (AUROC) scores close to unity, even in scenarios with low SNR \cite{gkelias2022gan}. 
}

In response to the complexity and training overhead of GANs and the low accuracy of traditional AE networks in anomaly detection of electromagnetic signals, the authors in \cite{cheng2022resnet} proposed a ResNet-AE network model. This model integrates the encoder and decoder with ResNet architecture and LSTM architectures for efficient feature mapping and data reconstruction. To process the anomaly detection results and establish an adaptive decision threshold, a K-Means classifier with two categories is constructed, using a random initial clustering center to categorize the anomaly scores. After iterative clustering, the centers for normal and abnormal signal scores are determined, and the mean value of these centers is used as the threshold for anomaly judgment.  When applied to radar signal anomaly detection, the proposed ResNet-AE method achieves a high recognition accuracy, exceeding 85\% \cite{cheng2022resnet}.

\ignore{
The original VAE is based on two loss functions: KL divergence and reconstruction error. These loss functions are also utilized as metrics for anomaly scoring. 
}
To further investigate the impact of different weightings of the KL divergence in the loss function of VAEs, the authors of \cite{harini2023data} proposed an approach for data anomaly detection using a $\beta$-VAE \cite{fil2021beta}. This advanced model, employing a multivariate normal distribution, introduces a coefficient $\beta$ to control the KL term. It allows for a more disentangled representation of data, where each unit in the latent code is responsive to a single generative element, enhancing the model's interpretability and effectiveness. However, the study does not specify the method for selecting the optimal value of the $\beta$ coefficient.
\ignore{
This aspect represents an area for further improvement and research, as the choice of $\beta$ can significantly influence the model's performance in anomaly detection tasks.
}

While existing methods have proven effective in anomaly detection, they often involve transmitting large volumes of raw data, resulting in significant channel interference and energy consumption. To address the substantial demand for computational resources, the study \cite{luo2018distributed} introduces an AE-based distributed anomaly detection approach in Wireless sensor network (WSN), characterized by its simplicity with only three layers. Each sensor in the proposed approach is equipped with a copy of the AE and is responsible for two primary tasks, in addition to its regular sensing function. The first task involves providing the input and output data of the AE to the IoT cloud, which serves as the training data. This data transfer occurs through a gateway or cluster head at a significantly lower frequency compared to the sensing rate. The second task is the execution of anomaly detection, which is conducted locally at the sensor level. This process is independent of any communication with other sensors, the gateway, or the IoT cloud, thereby enabling efficient and autonomous anomaly detection within each sensor unit, offering a more efficient and autonomous approach to anomaly detection in WSN \cite{luo2018distributed}.

Besides reconstruction-based methods, prediction-based methods have also proven to be effective for anomaly detection, which directly predict the probability of an anomaly without defining an anomaly score\cite{lu2021gan, zhou2021radio, toma2020ai, toma2020deep}. 

A primary challenge in physical layer sensing is the large amount of unclean and irrelevant data collected from sensors, known as data imbalance \cite{thabtah2020data}. This issue often results in traditional AI models misclassifying all samples as abnormal, further complicating anomaly detection \cite{lee2021gan}. To tackle the data imbalance problem, \cite{lu2021gan} introduces the MSGAN model, a GAN-based data augmentation strategy specifically designed for sensor anomaly detection. This model integrates WGAN-GP \cite{arjovsky2017wasserstein} with a novel adaptive update strategy during offline training. The adoption of an adaptive update strategy allows the MSGAN to accelerate training convergence and improve the quality of synthetic samples. 

\ignore{
In particular, simulation results demonstrate the significant impact of data imbalance on anomaly detection accuracy. To be more specific, with a 1:100 imbalance ratio, the accuracy of the LightGBM model is only 87.1\%. In contrast, by using the proposed model to increase the ratio to 1:1 leads to a marked improvement in accuracy, reaching 97.2\% \cite{lu2021gan}. 
}

In SAIFE \cite{rajendran2019unsupervised}, the distribution captured by AAEs corresponds more to the latent representations than to the original training samples. To address this limitation, the study in \cite{zhou2021radio} introduced an Encoder-GAN (E-GAN) structure, which incorporates an encoder network into the original GAN framework to reconstruct the spectrogram. By integrating an encoder into the standard GAN, the latent representations are controlled by the encoder rather than being randomly selected, which ensures that the generator produces data within the actual data distribution. Consequently, the E-GAN model is more adept at capturing the distribution of input samples than the SAIFE. However due to the convolutional structure in E-GAN, the time complexity of the proposed algorithm is higher than that of the SAIFE model where a fully-connected architecture is enough \cite{zhou2021radio}.

\ignore{
Regarding the network structures, a fully-connected architecture suffices for the discriminator in AAEs to process 1D input. However, a convolutional structure is more appropriate for the discriminator in E-GAN, which handles 2D spectrogram input. 

Therefore, the time complexity of the proposed algorithm is higher than that of the SAIFE model, indicating a trade-off between model complexity and performance \cite{zhou2021radio}.
}

In \cite{toma2020ai}, a framework integrating Dynamic Bayesian Networks (DBNs) \cite{ravanbakhsh2018learning} and GANS was proposed to detect abnormalities. A distinctive aspect of this approach is the use of a generalized state vector \cite{friston2014cognitive}, consisting of the signal feature extracted from the Stockwell Transform (ST) and the corresponding derivatives, as the input for the model. In the proposed framework, DBNs are used to learn switching models where each switching variable can be associated with a different linear dynamic model. This approach is particularly suited for scenarios involving low-dimensionality data due to the vocabulary size of switching variables. Conversely, CGAN is employed for scenarios involving high-dimensionality data. While GANs are capable of effectively managing a high number of different dynamic models implicitly, they have a notable limitation: unlike DBNs, GANs cannot manage uncertainty with probabilistic knowledge \cite{toma2020ai}.

\ignore{
While GANs are capable of effectively managing a high number of different dynamic models implicitly, they have a notable limitation: unlike DBNs, GANs cannot manage uncertainty with probabilistic knowledge. This distinction underscores the complementary strengths and limitations of these two AI-based techniques in addressing the challenges of abnormality detection in the physical layer security.
}

According the results in \cite{toma2020ai}, it is demonstrated that approaches utilizing generative learning of deep features yield superior results in anomaly detection when compared to conventional techniques, particularly the Cyclostationary Feature Detector (CFD) \cite{martian2014spectrum}. Therefore, the authors in \cite{toma2020deep} conducted a comparative analysis of three deep generative models: the CGAN, the ACGAN, and the VAE for spectrum anomaly detection in the millimeter Wave (mmWave) communications. Tested on a real dataset collect by The National Instruments mmWave Transceiver System \cite{toma2020deep}, all three models demonstrated commendable performance in anomaly detection, particularly the AC-GAN. The ROC curves from these tests confirmed that these models have a high probability of detection while maintaining a low false alarm rate. Furthermore, the VAE model demonstrates more efficient computational performance in both the training and testing processes compared to the other two networks. 
\ignore{
This efficiency is attributed to the simpler nature of the KL divergence used in VAE, as opposed to the mean-squared error approach employed in the other models.}

Similarly, the authors in \cite{rathinavel2022efficient} explored a range of generative model approaches, including U-Net WGAN, ResNet WGAN, and ResNet VAE applied to spectral density functions. The anomaly scoring mechanism employed varies with the model: binary cross-entropy loss is used between the input and reconstruction for U-Net WGAN and ResNet VAE, while mean-squared error loss is applied for ResNet WGAN. For comparison and validation, three well-known anomaly detection methods are used as baselines: Isolation Forest \cite{liu2008isolation}, One-class SVM \cite{wang2004anomaly}, and fAnoGAN \cite{schlegl2019f}. The results demonstrate excellent performance of these generative models compared to traditional baseline approaches for various types of anomalies. In particular, the Unet GAN achieves the highest average in four out of the five metrics \cite{rathinavel2022efficient}.

As summarized in Table \ref{tab:anomaly}, GAI models showcase superior performance in anomaly detection within complicated feature data than traditional AI models. However, some methods cannot be employed to characterize and classify the anomalous signals \cite{cheng2022resnet,toma2020deep}, which holds critical importance for the subsequent maintenance and security of network equipment. Consequently, future research should concentrate on creating advanced GAI models capable of detecting and classifying various anomalous signals.

\subsection{Data Reconstruction}

\begin{table*}[htp] \scriptsize
  \centering
  \caption{Summary of GAI for Data Reconstruction in Physical Layer \\Blue circles describe the methods; Green correct markers and Red cross markers represent pros and cons respectively.}
  \label{tab:data}
    \begin{tabular}{m{0.11\textwidth}<{\centering}||m{0.06\textwidth}<{\centering}|m{0.09\textwidth}<{\centering}|m{0.64\textwidth}<{\centering}}
      \hline
      \textbf{Techniques}  &  \textbf{Reference} & \textbf{Algorithm} & \textbf{Pros \& Cons} \\
       \hline
        \multirow{5}{0.11\textwidth}[-25pt]{\centering Data Reconstruction} & \cite{tran2018generative}  & SARGAN & \begin{itemize}[leftmargin=*]
        \item[\textcolor{blue-green}{\ding{108}}] A GAN network to recover this missing spectral information
        \item[\textcolor{green}{\ding{51}}] Not require any information of the missing band locations
        \item[\textcolor{green}{\ding{51}}] All computational complexity is at the training phase. 
        \item[\textcolor{red}{\ding{55}}] The requirement for extensive training data
        \vspace{-1.0em}
        \end{itemize}\\
        \cline{2-4}
    & \cite{feng2022waveform}  & VAE-GAN  & \begin{itemize}[leftmargin=*]
    \item[\textcolor{blue-green}{\ding{108}}] A VAE-GAN-based method for reconstructing DSSS signals
      \item[\textcolor{green}{\ding{51}}] Avoid complex parametric analysis of the signal
      \item[\textcolor{green}{\ding{51}}] Integrate DRSNs and self-attention in VAE-GAN
      \item[\textcolor{red}{\ding{55}}] Unsatisfactory effect in low SNR
      \vspace{-1.0em}
      \end{itemize}\\
      \cline{2-4}
    & \cite{guo2023high}  & MTS-GAN& \begin{itemize}[leftmargin=*]
    \item[\textcolor{blue-green}{\ding{108}}] A high-precision reconstruction method for electromagnetic environment data based on MTS-GAN
        \item[\textcolor{green}{\ding{51}}]  Use the GRUI to simulate time irregularities
      \item[\textcolor{green}{\ding{51}}]  High accuracy and convergence speed
      \item[\textcolor{red}{\ding{55}}] The specific requirements for training and implementing
      \vspace{-1.0em}
      \end{itemize}\\
      \cline{2-4}
            &  \cite{estiri2020variational} & VAE & \begin{itemize}[leftmargin=*]
            \item[\textcolor{blue-green}{\ding{108}}]  Investigate the performance of VAEs and compare the results with standard AEs
          \item[\textcolor{green}{\ding{51}}]  Use SSIM metric instead of the peak signal-to-noise ratio
      \item[\textcolor{red}{\ding{55}}] Limited in terms of the variety of noise models
      \vspace{-1.0em}
      \end{itemize}\\
      \cline{2-4}
     & \cite{wu2023cddm}  & CDDM & \begin{itemize}[leftmargin=*]
     \item[\textcolor{blue-green}{\ding{108}}] A channel denoising diffusion models for wireless communications to eliminate the channel noise
          \item[\textcolor{green}{\ding{51}}] Eliminate the channel nosie under Rayleigh fading channel and AWGN channel
      \item[\textcolor{red}{\ding{55}}] Relatively long sampling time
      \vspace{-1.0em}
      \end{itemize}\\
      \hline
    \end{tabular}
\end{table*}


Data reconstruction focuses on retrieving the original signal or information from corrupted or incomplete datasets \cite{chai2020deep}. This process involves various techniques to restore or approximate the original data, aiming to overcome the issues caused by interference and noise. 

\ignore{Data augmentation, on the other hand, is centered on generating additional usable data from the limited existing dataset. It involves applying transformations or manipulations to enhance the features of the existing data, thereby expanding the dataset. This expansion is crucial for improving data analysis and model training, particularly in scenarios where the amount of collected data is inherently limited.}

Traditional reconstruction methods, based on sparse representation and low-rank matrix completion \cite{nguyen2014sparse}, assume that both full-spectrum data and their corrupted counterparts are sparsely represented with a full-spectrum and a gapped-spectrum dictionary, respectively. Therefore, both representations are similarly sparse and share identical sparse codes. Consequently, these reconstruction methods lack the ability and robustness to distinguish closely situated targets at high resolution accurately without prior knowledge of the missing frequency bands. However, GAI models excel in learning complex data patterns, effectively reducing the dependency on prior knowledge of missing frequency bands. Moreover, GAI models leverage their advanced learning capabilities to data gaps, offering a more robust and flexible approach to signal reconstruction.

\ignore{
GAI models are particularly adept at generating and reconstructing data, effectively reducing the dependency on prior knowledge of missing frequency bands. Moreover, GAI models leverage their advanced learning capabilities to data gaps, offering a more robust and flexible approach to signal reconstruction.}

In \cite{tran2018generative}, the authors introduced a GAN framework named SARGAN, designed to reconstruct missing spectral information in Ultra-wideband (UWB) radar systems across multiple frequency bands. Specifically, SARGAN focuses on recovering Synthetic Aperture Radar (SAR) data \cite{moreira2013tutorial}. To train the GAN model, the model uses numerous data pairs, each comprising an uncorrupted scene and its frequency-corrupted version. The corrupted datasets are simulated by removing random frequency bands from the original data. A significant advantage over conventional spectral recovery methods is that the proposed model does not need any prior knowledge of the missing data. This is particularly beneficial in unpredictable scenarios including battlefield conditions, where jamming and interference can occur unexpectedly. The simulation results show that the recovered signals using SARGAN achieve an average gain of over 18 dB in SNR, even when up to 90\% of the operating spectrum is missing.

Compared to radar data, DSSS signals possess more complex structures which makes it challenging to characterize accurately the properties of a target signal. To extract more properties such as Pseudonoise (PN) sequence \cite{helleseth2021pseudo}, a method based on VAE-GAN \cite{larsen2016autoencoding} for reconstructing DSSS signals was proposed in \cite{feng2022waveform}. By integrating VAE and GAN, the encoder provides the generator with a loss function that measures the discrepancy between real and generated data. Furthermore, the proposed framework incorporates a Deep Residual Shrinkage Network (DRSN) \cite{jiang2022image} and a self-attention mechanism \cite{vaswani2017attention} into the encoder and discriminator. The DRSNs are effective in minimizing redundant information in the collected signal, particularly noise-induced redundancy. Meanwhile, the self-attention mechanism facilitates the establishment of long-distance dependencies within the input sequences. However, while the proposed model is adaptable to PN sequences with varying code lengths, its performance in low SNR environments significantly diminishes. Particularly, when the SNR falls below 13 dB, there is a sharp decline in the model's performance \cite{feng2022waveform}.

\begin{figure}[htp]
    \centering
    \includegraphics[width= 1.0\linewidth]{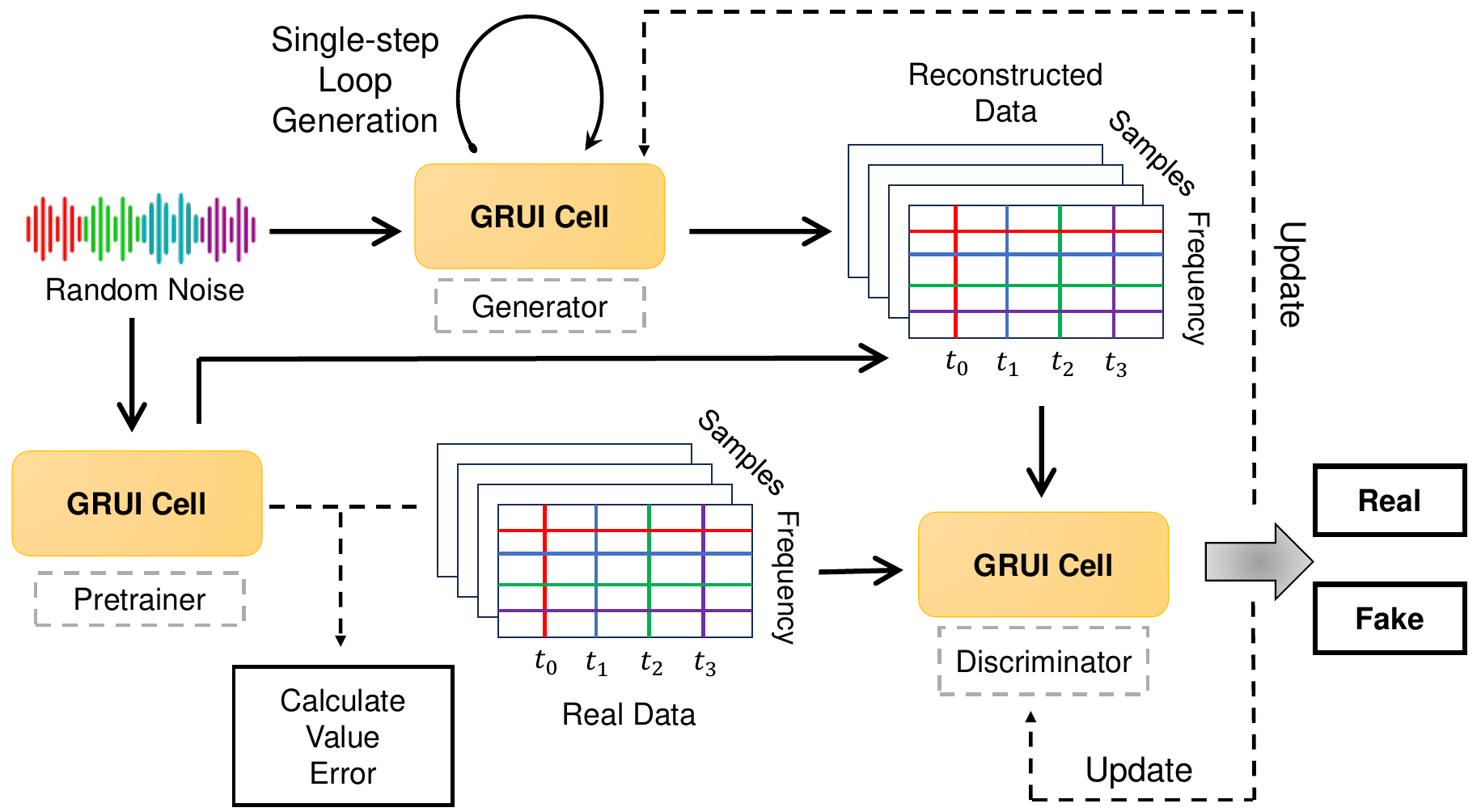}
    \caption{The MTS-GAN data completion network structure \cite{guo2023high}. The generator network is built by the gated loop unit GRUI for data interpolation. GRUI can simulate time irregularities allowing for more accurate extraction of the distribution characteristics of time-frequency signal data.}
    \label{fig:constru}
\end{figure}

To improve the precision of reconstructing electromagnetic environment data, the authors in \cite{guo2023high} developed a high-precision method using a Multi-Component Time Series Generation Adversarial Network (MTS-GAN). This approach effectively utilizes multivariate time series data to better capture the correlations between the time and frequency domains of electromagnetic data (Fig. \ref{fig:constru}). A key component of this method is the use of a Gate Recursive Unit (GRUI), which simulates time irregularities. The GRUI allows for more accurate extraction of the distribution characteristics of time-frequency signal data and reduces the impact of random losses in time series. The proposed method achieves high accuracy and ensures rapid convergence and iterative optimization speeds with the 'Qingdao Offshore Measurement Data Set' \cite{guo2023high}.

Besides GANs, VAEs are recognized as another powerful GAI model for effective data reconstruction. In \cite{estiri2020variational}, a VAE-based model was proposed for data construction in noisy channels. The AE and VAE models are particularly effective in regularizing the latent space distribution, a feature that is highly beneficial in data reconstruction with Gaussian noise channels.
\ignore{
However, traditional AEs \cite{bourtsoulatze2019deep} perform poorly, particularly under low SNR conditions. In such environments, data points in the latent space can undergo random deviations from their original positions. In contrast, 
}
The VAE's regularized latent space facilitates accurate decoding by the decoder, thus improving performance in noisy settings. Consequently, when evaluated on the STL10 dataset \cite{coates2011analysis} using the Structural Similarity Index (SSIM) metric \cite{brunet2011mathematical}, the VAE-based method demonstrates a smoother output attributed to its inherent structure.

DMs have garnered attention in the field of wireless communications due to their inherent ability to progressively remove noise, especially in aiding receivers to mitigate channel noise. In response to this potential, the study \cite{wu2023cddm} introduced Channel Denoising Diffusion Models (CDDM) specifically designed for wireless communications. CDDM aims to leverage the noise reduction properties of DMs to enhance the quality and reliability of signal reception in wireless communication channels. CDDM is trained using a specialized noise schedule specifically adapted to the wireless channel, enabling the effective elimination of channel noise through sampling algorithm. The training algorithm for the combined CDDM and JSCC system is structured into three distinct stages. In the first and last stages, the JSCC encoder and decoder are trained to minimize the reconstruction error. The second stage involves fixing the parameters of the JSCC encoder, thereby allowing the CDDM to learn the distribution of latent representations. This stage utilizes a noise schedule that closely simulates the distribution of channel noise, rendering the CDDM adaptable to a variety of channel conditions. The results demonstrate that systems incorporating CDDM consistently outperform those without CDDM across all SNR regimes, under both AWGN and Rayleigh fading channels. Notably, under an AWGN channel and a Rayleigh fading channel at 20 dB SNR, the CDDM achieves 0.49 dB gain and 1.06 dB gain, respectively \cite{wu2023cddm}.

The applications of GAI for data reconstruction in Table \ref{tab:data} showcase its remarkable ability to process and regenerate missing or corrupted data, ensuring communication integrity. However, the performance is still limited in low SNR scenarios \cite{feng2022waveform}. Therefore, proposing more accurate models in high noise situations is a future direction.

\section{Future Research Direction}\label{OI}

Despite its impressive capabilities in complex data feature extraction, reconstruction, and enhancement, the applications of GAI in physical layer security are still in its early stages. This section aims to explore the open issues and research directions related to the integration of GAI in physical layer security.

\subsection{Model Improvements}

Enhancing physical layer security necessitates models that significantly advance in terms of robustness and efficiency requiring model improvements. By incorporating advanced neural network architectures, GAI systems can learn to simulate and counteract time related attack patterns more effectively \cite{jiang2021transgan}. Enhancements in adversarial training techniques will also enable GAI models to better mitigate potential vulnerabilities \cite{boppana2023gan}.
\ignore{
Enhancements in adversarial training techniques will also enable these models to better anticipate and mitigate potential vulnerabilities, ensuring a higher level of security in communications \cite{boppana2023gan}. 
}
Moreover, GAI-aided information encryption may be further explored in conjunction with the near-field beam focusing via Extremely large-scale multiple-input-multipleoutput (XL-MIMO) that exploits the propagation characteristics of both distance and direction. The latter enables to focus the transmitted signal energy onto an intended user, so as not to induce information leakage to eavesdroppers. This certainly enhances the physical layer security for emerging 6G Wireless equipped with XL-MIMO \cite{wang2024tutorial}.



\subsection{Multi-scenario Deployment}

As the deployment of GAI in physical layer security, its application across various scenarios emerges as a critical area of focus. The intricate architecture of GAI poses challenges for its implementation on edge devices, often requiring the transmission of additional data \cite{luo2018distributed}. Incorporating distributed deployment strategies, GAI can efficiently leverage edge computing capabilities, thus minimizing latency and reducing the need for extensive data transmission by processing information closer to its source \cite{du2023exploring}. Furthermore, the Mixture of Experts (MoE) model can dynamically assign tasks to specialized sub-models or ‘experts’ \cite{shi2019variational}. It presents a promising avenue for enhancing the adaptability and efficiency of GAI in addressing the multifaceted and intricate scenarios encountered in physical layer security. Exploring the integration of the MoE model with GAI to leverage the strengths of both approaches is a noteworthy direction for future research.

\ignore{
\subsection{Model Privacy Protection}

In deep learning, the rise of model-targeted attacks has highlighted the critical need for enhanced defensive measures \cite{abadi2016deep}. These malicious efforts seek to steal model training data and parameters, enabling attackers to devise more sophisticated and targeted attacks. The extended training times and complex architectures associated with GAI increase their susceptibility, making it challenging to implement swift counteractions after a security breach \cite{jordon2018pate}. This situation has prompted a move towards incorporating privacy-preserving strategies, particularly differential privacy \cite{dockhorn2022differentially}, into model development to prevent data and parameter exposure. However, there remains seldom in integrating differential privacy with algorithms aimed at physical layer security detection. The initiative to merge differential privacy with physical layer security methods marks a crucial step towards fortifying AI defenses against the continuously advancing threats in the cyber landscape. The initiative to merge differential privacy with physical layer security methods marks a crucial step towards fortifying AI defenses against the continuously advancing threats in the cyber landscape.
}

\subsection{Resource-Efficient Optimization}

Compared with traditional AI, GAI usually requires more resources for training and inference due to its complex mission objectives. It causes serious burden and impact on the normal process operation of the device, especially for devices with limited resources such as mobile phones. Therefore, future directions should emphasize the development of lightweight GAI models that can operate with minimal computational resources while maintaining high security standards \cite{chen2020animegan}. For instance, adapting model pruning techniques to remove unnecessary parameters from GANs without compromising their ability to generate or discriminate can significantly reduce the computational load \cite{vo2022ppcd}. Additionally, exploring federated learning approaches could decentralize the training process, allowing GAIs to learn from diverse datasets across multiple devices while ensuring data privacy and reducing the need for centralized, powerful computing resources \cite{du2023beyond}. These strategies promise to enhance the scalability of GAIs in securing the physical layer and ensure their applicability in resource-constrained environments including IoT devices and edge computing platforms, where security and efficiency are paramount.

\subsection{Secure SemCom}

"GAI-aided Secure SemCom" is certainly a vital future research direction. The task-oriented SemCom aims at minimizing the transmission overhead in resource-constrained networks, such as AI-native wireless networks \cite{liang2023generative}. Additionally, it focuses on performing a given task properly with the aid of GAI as well as knowledge base at both ends, even though the reconstructed data is not exactly same as the original data \cite{du2023semantic}. Consequently, the performance metric transitions from bit-level accuracy including BER, to the degree of task fulfillment within a specified QoE value, given the GAI with knowledge base is shared between the transceiver. Therefore, this paradigm shift necessitates a reevaluation of GAI model design criteria within SemCom, focusing on task fulfillment levels facilitated by the synergistic use of GAI and a shared knowledge base in physical layer security \cite{du2023rethinking}.

\section{Conclusion}\label{conclu}

\ignore{
GAI emerges as a promising technology for enhancing physical layer security, attributed to its proficiency in complex data feature extraction, reconstruction, and enhancement. This article presents a comprehensive survey on the applications of GAI in physical layer security.}

This paper has presented a comprehensive survey on the applications of GAI in physical layer security, attributed to its remarkable capabilities in extracting, reconstructing, and enhancing complex data features. It introduced the background of GAI, encompassing its architecture, classification, and foundational models. Subsequently, it explored various security properties such as communication confidentiality, authentication, availability, resilience, and integrity. Finally, it highlighted crucial future research directions for generative AI in physical layer security, which underscores the potential of GAI to further enhance security measures, demonstrating its vital role in safeguarding communication networks against evolving security threats.

\ignore{GAI has emerged as a promising technology for enhancing physical layer security, attributed to its remarkable capabilities in extracting, reconstructing, and enhancing complex data features. This article provided a comprehensive survey that explores the wide-ranging applications of GAI in physical layer security. Initially, we introduced the background of GAI, encompassing its architecture, classification, and foundational models. Subsequently, we offer detailed reviews on data reconstruction and augmentation, illustrating how GAI can significantly contribute to the robustness of physical layer security. Furthermore, we explored various security properties such as communication confidentiality, authentication, availability, resilience, and integrity. This includes discussions on mitigating and defending against typical attacks encountered in physical layers, and reconstruct data received by receivers to ensure data integrity. Additionally, this article underscored the importance of identifying crucial future research directions for generative AI in physical layer security. By emphasizing the potential of GAI to strengthen security measures, it highlights the vital role it plays in safeguarding communication networks against the ever-evolving security threats. }

\ignore{
Concluding, we highlight crucial future research directions for generative AI in physical layer security. This underscores the potential of GAI to further enhance security measures, demonstrating its vital role in safeguarding communication networks against evolving security threats.
}

\bibliographystyle{IEEEtran}
\bibliography{Ref}

\end{document}